\documentclass[journal]{IEEEtran}
\IEEEoverridecommandlockouts
\usepackage{cite}
\usepackage{amsmath,amssymb,amsfonts}
\usepackage{graphicx}
\usepackage{subcaption}
\usepackage{booktabs}
\usepackage{textcomp}
\usepackage{xcolor}
\usepackage{tikz}
\usetikzlibrary{shapes.geometric, arrows.meta, positioning, fit, backgrounds}
\usepackage{amsthm}
\usepackage[moderate]{savetrees}
\newtheorem{proposition}{Proposition}
\newtheorem{remark}{Remark}
\newtheorem{theorem}{Theorem}
\newtheorem{lemma}{Lemma}
\newtheorem{cor}{Corollary}
\newtheorem{definition}{Definition}
\DeclareMathOperator{\sinc}{sinc}

\def\BibTeX{{\rm B\kern-.05em{\sc i\kern-.025em b}\kern-.08em
 T\kern-.1667em\lower.7ex\hbox{E}\kern-.125emX}}

\definecolor{analogblue}{RGB}{180,210,240}
\definecolor{digitalorg}{RGB}{255,210,160}
\definecolor{boundarycolor}{RGB}{220,220,220}

\title{Time-Based Sampling in Shift-Invariant Spaces
\thanks{This work was supported in part by the European Research Council (ERC)
under the European Union's Horizon 2020 Research and Innovation program
(grant No.\ 101000967, 646804-ERC-COG-BNYQ), by the European Union
(ERC, SWIMS, 101119062), and by the Israel Science Foundation under Grant
0100101.}}

\author{\IEEEauthorblockN{Hila~Naaman$^{\dagger}$,~\IEEEmembership{Member,~IEEE,}
 Yhonatan~Kvich$^{\ast}$,~\IEEEmembership{Graduate~Student~Member,~IEEE,}
 and~Yonina~C.~Eldar$^{\ast{\ddagger}}$,~\IEEEmembership{Fellow,~IEEE}}\\
\IEEEauthorblockA{$^{\dagger}$Department of Computer Science,
The College of Management Academic Studies, Israel\\
$^{\ast}$Faculty of Mathematics and Computer Science,
Weizmann Institute of Science, Israel\\
{$^{\ddagger}$Department of Electrical and Computer Engineering,
Northeastern University, Boston, MA 02115 USA}\\
hilana2@colman.ac.il, \{yonatan.kvich, yonina.eldar\}@weizmann.ac.il}}

\begin{document}
\maketitle

\begin{abstract}
Time encoding replaces the clocked amplitude sampler of an analog-to-digital converter by an asynchronous circuit that records only firing instants. Shift-invariant (SI) signals carry a fixed number of coefficients per period, yet guarantees for inverting firing times directly tie the event density to the generator bandwidth. We recover SI signals from integrate-and-fire time-encoding machine (IF-TEM) firing times at a rate set by the degrees of freedom. An analog front end gives each encoder a band-limited input whose Nyquist rate equals this rate. Low-pass filtering may remove the Riesz property of the generator shifts. A matched prefilter and a periodic mixer give stable recovery when the Riesz sum of the original generator over the mixer harmonics is bounded away from zero, a bound approaching the Riesz bound under continuity conditions. We give exact recovery and stability conditions in the encoder bias, gain and threshold, extend them to several generators, where distinct mixers are necessary, and derive an asymptotic per-coefficient mean-squared error bound for interval quantization, decreasing by six decibels per bit. Simulations confirm recovery near this rate despite in-band zeros, and, for a band-pass generator under timing error, 25 dB lower coefficient error than direct SI inversion at equal event counts.
\end{abstract}

\begin{IEEEkeywords}
Time encoding machine, integrate-and-fire, sub-Nyquist sampling,
shift-invariant spaces, time-based ADC, quantization, multiple generators.
\end{IEEEkeywords}


\section{Introduction}
\label{sec:intro}

Analog-to-digital conversion typically combines clocked amplitude sampling with quantization~\cite{nyquist1928certain,shannon1949communication,eldar2015sampling}. Clock distribution and amplitude conversion become increasingly costly in power as the sampling rate and required resolution grow~\cite{gray1998quantization,mulleti2024power}. Time-encoding machines (TEMs) instead represent a signal by event times~\cite{miskowicz2018event,camera1,neuromorphic_computer}. The integrate-and-fire TEM (IF-TEM) records an event whenever the integral of its biased input reaches a prescribed threshold~\cite{lazar2004perfect,lazar2004time,alvarado2011integrate,feichtinger2012approximate}. Its bias, gain, and threshold control the event density, while timing circuits determine the measurement resolution. This architecture shifts the acquisition task from clocked amplitude readout to asynchronous timing, and hardware implementations have been demonstrated~\cite{rastogi2011integrate,ryu2021time,naaman2023hardware}.

We consider signals in shift-invariant (SI) spaces, generated by linear combinations of known functions shifted on a uniform grid~\cite{unser2000sampling,aldroubi2001nonuniform,deboor1994structure,eldar2015sampling}. The unknowns are the expansion coefficients. One coefficient per period per generator determines the signal, independently of the generator bandwidths. We call this number of unknown coefficients per unit time the coefficient rate and use it as the reference for all firing rates in this paper. Under the Riesz condition and without additional structure such as sparsity, the coefficient rate is the minimal total rate of classical uniform SI sampling~\cite{eldar2009compressed}. Matched sampling filters with digital correction attain this rate~\cite{eldar2015sampling,christensen2004oblique}. Uniform sampling folds the spectral copies of each filtered generator into one frequency interval. Consequently, zeros in one copy need not prevent recovery: the Riesz condition bounds the sum of the squared magnitudes of all copies away from zero.

One line of work recovers SI signals directly from their events. Gontier and Vetterli~\cite{gontier2014sampling} established reconstruction from both crossing and integrate-and-fire events. Their sufficient density condition depends on the high-frequency content of the generator, rather than only on its coefficient rate. This distinction is important for band-pass generators, whose carrier frequencies can be much higher than their coefficient rates. Extensions address several generators observed by one encoder~\cite{soundarapandian2024multigenerator} and recovery of band-limited (BL) and spline signals from neuromorphic events~\cite{kamath2023neuromorphic}. An alternative approach preprocesses the signal into a class with an established decoder. Sampling kernels enable this reduction for pulse streams encoded by IF-TEMs~\cite{naaman2021fri,alexandru2019reconstructing}, and low-pass preprocessing is used in modulo sampling of SI signals~\cite{kvich2024modulo}.

Here we map each encoder input to the band whose Nyquist rate equals the per-generator coefficient rate. This band-limiting is our design choice: it lets us reuse BL time-encoding recovery and express a sufficient density condition directly in the IF-TEM parameters. With a low-pass filter (LPF) alone, the coefficients are identifiable when the generator is nonzero almost everywhere on the retained band; stable inversion additionally requires its magnitude to be bounded away from zero. The Riesz condition on the original generator does not ensure this baseband condition, because the LPF discards the other spectral copies.

An analog periodic mixer before the LPF restores spectral folding in the analog domain, as in the modulated wideband converter~\cite{mishali2010theory} and modulo sampling of SI signals~\cite{kvich2024modulo}. Its harmonics select and weight the copies brought into the encoder band. With a matched prefilter, equal mixer weights produce the Riesz sum of the original generator truncated to the selected harmonics. Recovery is stable when this truncated sum has a positive lower bound. Under the continuity assumptions stated below, that bound approaches the optimal lower Riesz bound as harmonics are added. Compared with direct SI inversion, the proposed architecture adds an analog front end: an LPF in each channel, a mixer when spectral folding is needed, and a wideband matched prefilter for the truncated-Riesz-sum construction.

The contributions are threefold. First, we establish single-generator recovery and stability conditions, beginning with a linear time-invariant (LTI) filter and then adding a mixer when the baseband response is insufficient. The density condition is inherited from BL recovery; the effective generator response determines the conditioning of the SI inversion. Second, we extend the construction to multiple generators. Within the common-baseband architecture, LTI filters alone give a rank-one response. Distinct mixers enable separation: full column rank almost everywhere ensures identifiability, whereas stability requires a uniform positive lower bound on the smallest singular value. With one channel per generator, the total event rate can approach the coefficient rate.

Third, we derive an asymptotic per-coefficient mean-squared error (MSE) bound for stochastic interval quantization, with six-decibels-per-bit decay. Its prefactor depends on the encoder operating point and the conditioning of the analog response. We distinguish comparisons at equal bits per measurement from those at equal total bit rate. Numerical experiments evaluate both the designed-mixer and matched-prefilter architectures. For a band-pass generator at equal event counts and under timing error, the matched front end lowers the coefficient error by about 25~dB relative to direct SI inversion of IF-TEM firing times based on the measurement model of Gontier and Vetterli~\cite{gontier2014sampling}. Direct inversion is slightly more accurate for the cubic spline, by $0.7$--$1.0$~dB at the reported jitter levels. A separate comparison with crossing-based encoding uses a common BL input to isolate the encoders.

Section~\ref{sec:background} reviews BL time encoding and classical SI sampling with filters and mixers. Section~\ref{sec:SIrec} develops single-generator recovery, the mixer construction, and the multigenerator extension. Sections~\ref{sec:quant} and~\ref{sec:sim} present the quantization analysis and numerical experiments, respectively. Section~\ref{sec:conc} concludes the paper. Proofs and additional simulations are given in the supplementary material.

\section{Background and Problem Formulation}
\label{sec:background}
 \emph{Notation.} Throughout the paper, the continuous-time Fourier
transform (CTFT) of a signal $g(t)\in L^2(\mathbb{R})$ is denoted by
$G(\omega)=\int_{-\infty}^{\infty} g(t)\,e^{-j\omega t}\,dt$, and the
discrete-time Fourier transform (DTFT) of a sequence $a[n]\in\ell^2$ by
$A(e^{j\omega T})=\sum_{n\in\mathbb{Z}} a[n]\,e^{-j\omega n T}$, which is
$2\pi/T$-periodic in $\omega$. With this convention, the CTFT of a signal
$x(t)=\sum_n a[n]\,h(t-nT)$ factors as $X(\omega)=A(e^{j\omega T})\,H(\omega)$.
All signals are real valued. For Hermitian matrices $\mathbf{A},\mathbf{B}$,
the L\"owner order $\mathbf{A}\preceq\mathbf{B}$ means that $\mathbf{B}-\mathbf{A}$
is positive semidefinite.

\subsection{Integrate-and-Fire Time-Encoding Machine}
\label{subsec:IAF}

\begin{figure}[t!]
 \centering
 \includegraphics[width=2.5in]{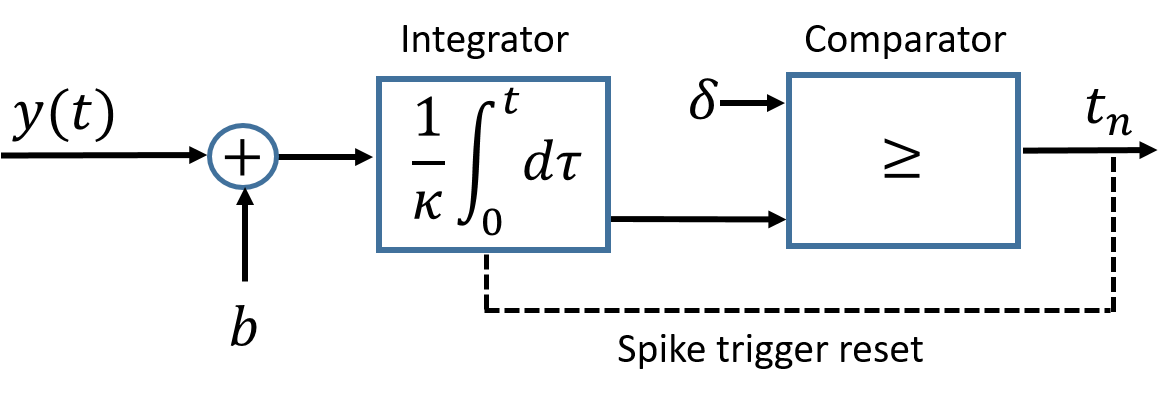}
 \caption{Integrate-and-fire time-encoding machine. The input is biased
 by $b$, scaled by $\kappa$, and integrated. A spike is recorded whenever
 the integrator reaches the threshold $\delta$, after which the integrator
 resets.}
 \label{fig:Vetterli_TEM}
\end{figure}

We employ an IF-TEM following the architecture of~\cite{lazar2004time}, assuming
a zero-length refractory period (Fig.~\ref{fig:Vetterli_TEM}). The IF-TEM maps
a bounded analog input $y(t)$ to a sequence of spike times
$\{t_n,\,n\in\mathbb{Z}\}$, defined by three positive constants: the bias $b$,
the scaling factor $\kappa$, and the threshold $\delta$.

A bias $b$ is added to $y(t)$, which is $c$-bounded with $|y(t)| \leq c < b$.
The sum is integrated and scaled by $1/\kappa$. Whenever the integrator reaches
$\delta$, a spike time $t_{n+1}$ is recorded and the integrator resets. For any
pair of consecutive firing times,
\begin{equation}
 \frac{1}{\kappa}\int_{t_n}^{t_{n+1}} \bigl(y(s)+b\bigr)\,ds = \delta.
 \label{eq:fundamental_relation}
\end{equation}
The integral measurements
\begin{equation}
 y_n \triangleq \int_{t_n}^{t_{n+1}} y(s)\,ds = \kappa\delta - b(t_{n+1}-t_n)
 \label{eq:measurement_y_n}
\end{equation}
are computable directly from the spike times and the known hardware parameters.
Since $|y(t)| \leq c$, the inter-spike intervals satisfy
\begin{equation}
 \frac{\kappa\delta}{b+c} \leq t_{n+1}-t_n \leq \frac{\kappa\delta}{b-c},
 \label{eq:interval_bounds}
\end{equation}
so the firing rate $F_R$ is bounded by
\begin{equation}
 \frac{b-c}{\kappa\delta} \leq F_R \leq \frac{b+c}{\kappa\delta}.
 \label{eq:firing_rate_limits}
\end{equation}

\subsection{Time-Based Reconstruction of BL Signals}
\label{subsec:rec_IF}

Recovery of signals from IF-TEM measurements is well established for
$c$-bounded, $2\Omega$-BL inputs~\cite{lazar2004perfect,lazar2004time,adam2020sampling,naaman2021time}.

\begin{definition}
\label{def:bl}
A signal $y(t) \in L^2(\mathbb{R})$ is \emph{$c$-bounded and $2\Omega$-BL}
if $|y(t)| \leq c$ for all $t$ and its Fourier transform is supported on
$[-\Omega, \Omega]$.
\end{definition}

By the Nyquist-Shannon theorem, such a signal can be perfectly reconstructed
from uniform samples at rate $\Omega/\pi$~\cite{nyquist1928certain}. For the
IF-TEM, perfect recovery is achievable provided $b > c$ and the maximum
inter-spike interval satisfies
\begin{equation}
 \Delta t_{\max} = \frac{\kappa\delta}{b-c} < \frac{\pi}{\Omega}.
 \label{eq:density}
\end{equation}
Equation~\eqref{eq:density} requires the IF-TEM to fire above the Nyquist rate.
The reconstruction operator
\begin{equation}
 \mathcal{R}(y)(t) = \sum_{i\in\mathbb{Z}}
 y_i\,\mathrm{sinc}_\Omega(t - s_i),
 \label{eq:rec_operator}
\end{equation}
where $s_i = (t_i+t_{i+1})/2$ and
\begin{equation}
 \mathrm{sinc}_\Omega(t) =
 \begin{cases}
 \dfrac{\sin(\Omega t)}{\pi t}, & t \neq 0,\\[4pt]
 \dfrac{\Omega}{\pi}, & t = 0,
 \end{cases}
 \label{eq:sinc_def}
\end{equation}
is applied iteratively~\cite{lazar2004perfect}:
\begin{equation}
 y_{l+1}(t) = y_l(t) + \mathcal{R}\bigl(y(t) - y_l(t)\bigr),
 \quad y_0 = \mathcal{R}(y).
 \label{eq:iteration}
\end{equation}
The iteration converges, $\lim_{l\to\infty}y_l(t)=y(t)$, with error
\begin{equation}
 \|y - y_l\|_{L^2} \leq r^{l+1}\|y\|_{L^2},
 \quad r = \frac{\kappa\delta\Omega}{(b-c)\pi} < 1.
 \label{eq:convergence}
\end{equation}
The operator $\mathcal{R}$ and the iteration \eqref{eq:iteration} form the
inner stage of the framework developed in this paper, converting the firing
times into the BL signal seen by the encoder. Our task is to design the analog
front end and the digital correction so that this BL signal determines the SI
coefficients.

\subsection{Shift-Invariant Spaces}
\label{subsec:si}

SI spaces describe signals expressed as linear combinations of shifted
generator functions on a fixed grid~\cite{eldar2015sampling}. An SI space with a
single generator $h(t)$ is
\begin{equation}
 \mathcal{V}:= \left\{
 \sum_{n\in\mathbb{Z}} a[n]\,h(t-nT)
 \;\middle|\;
 a \in \ell^2
 \right\}.
 \label{eq:si}
\end{equation}
Choosing $h(t) = \mathrm{sinc}(\pi t/T)$ recovers the space of
$2\Omega$-BL signals with $\Omega = \pi/T$, so BL signals
are a special case of the SI framework.

Signals in $\mathcal{V}$ are perfectly recovered from uniform samples at rate
$T^{-1}$, whether or not the generator or the signal is
BL~\cite{eldar2015sampling}. The signal is passed through an analog prefilter
$s(t)$ and sampled at times $nT$, producing samples
$c[n]=\bigl(s(-\cdot)*x\bigr)(nT)$ whose DTFT satisfies
$C(e^{j\omega T})=R_{SH}(e^{j\omega T})\,A(e^{j\omega T})$, where
\begin{equation}
 R_{SH}(e^{j\omega T})
 = \frac{1}{T}
 \sum_{k\in\mathbb{Z}}
 \overline{S\!\left(\omega-\tfrac{2\pi k}{T}\right)}
 H\!\left(\omega-\tfrac{2\pi k}{T}\right),
 \label{eq:rsh}
\end{equation}
$S(\omega)$ and $H(\omega)$ are the Fourier transforms of $s(t)$ and
$h(t)$, and $R_{SH}(e^{j\omega T})$ is the DTFT of
$r_{sh}[n] = (s(-\cdot)*h(\cdot))|_{t=nT}$.
The coefficients are recovered by applying the digital \emph{correction filter}
$1/R_{SH}(e^{j\omega T})$ to the samples $c[n]$, and stability requires
$R_{SH}(e^{j\omega T})$ to be bounded away from zero and infinity on the band,
a Riesz-type condition on the pair
$(s,h)$~\cite{eldar2015sampling,christensen2004oblique} that reduces for $s=h$
to the Riesz basis condition \eqref{eq:riesz_single} below. Two features of
\eqref{eq:rsh} are used repeatedly in what follows. First, all spectral copies
of $H$ contribute to the sum, and the mechanism producing it is a modulation:
sampling on the grid $\{nT\}$ is multiplication by the $T$-periodic Dirac comb
$\sum_n\delta(t-nT)$, which folds the copies $H(\omega-2\pi k/T)$ into
\eqref{eq:rsh}. The time-encoding pipeline will have to reproduce that folding
by other means (Section~\ref{subsec:robustness}). Second, no band-limiting
appears anywhere in this chain, and the rate remains $1/T$ however wideband
$h(t)$ is, since it is the single degree of freedom per period, and not the
bandwidth, that fixes it.

SI spaces can be extended to $N$ generators $\{h_l(t)\}_{l=1}^N$:
\begin{equation}
 \mathcal{V}:= \left\{
 \sum_{l=1}^{N}\sum_{n\in\mathbb{Z}} a_l[n]\,h_l(t-nT)
 \;\middle|\; a_l \in \ell^2,\;\forall l
 \right\}.
 \label{eq:si_multi}
\end{equation}
The samples are acquired through a bank of $N$ sampling filters
$s_i(t)$, one per generator~\cite{eldar2015sampling}, and recovery inverts the
$N\times N$ matrix $\mathbf{M}_{SH}(e^{j\omega T})$ whose $(i,l)$-th entry is
$R_{S_iH_l}(e^{j\omega T})$, defined as in \eqref{eq:rsh}; stability requires
this matrix to be boundedly invertible for almost every $\omega$. For the
matched choice $s_i=h_i$ it is the matrix $\mathbf{M}_{HH}(e^{j\omega T})$ with
entries $R_{H_iH_l}(e^{j\omega T})$, and the requirement becomes the
multi-generator Riesz condition \eqref{eq:riesz_multi} below.
An alternative classical
architecture replaces the filter bank by a bank of periodic modulators
(mixers), as in the MWC~\cite{mishali2010theory,mishali2011subnyquist} and, for
SI spaces seen through a nonlinear front end, in modulo sampling of SI
signals~\cite{kvich2024modulo}; we return to this parallel in
Section~\ref{subsec:multi}, where the
time-encoding setting singles out the mixer architecture.

SI spaces are related to, but distinct from, the finite-rate-of-innovation
(FRI) models treated in the time-encoding
literature~\cite{naaman2021fri,alexandru2019reconstructing}. There the pulses
sit at unknown, off-grid locations, so that recovery requires estimating delays,
a nonlinear problem sensitive to noise and dependent on the model order. Here
the shift grid is known and only the linear coefficients are unknown, so that
recovery is linear and stable, while the signal may still have infinitely many
degrees of freedom overall.

Stability of SI recovery follows from Riesz basis conditions. For a single
generator, $\{h(t-nT)\}_{n\in\mathbb{Z}}$ forms a Riesz basis if and only if
there exist $0 < \alpha_0 \leq \beta_0 < \infty$ such that
\begin{equation}
 \alpha_0 \leq R_{HH}(e^{j\omega T}) \leq \beta_0
 \quad \text{for almost all } \omega,
 \label{eq:riesz_single}
\end{equation}
where $R_{HH}(e^{j\omega T}) = \frac{1}{T}\sum_{k}|H(\omega-2\pi k/T)|^2$.
When $h(t)$ is baseband-confined to $|\omega|\leq\pi/T$, a single term
contributes to the sum on the band and \eqref{eq:riesz_single} is equivalent to
$\sqrt{T\alpha_0} \leq |H(\omega)| \leq \sqrt{T\beta_0}$ a.e.\ on
$|\omega| \leq \pi/T$. For a wideband generator, however, several copies
contribute to $R_{HH}$, so that \eqref{eq:riesz_single} may hold while the
baseband copy $H(\omega)$ itself vanishes at some
$\omega\in[-\pi/T,\pi/T]$, the remaining copies keeping the sum positive. The
Riesz condition is thus a condition on the aliased sum, and not on the single
baseband copy. This distinction is immaterial for uniform SI sampling, whose
sampler folds all copies and, with the matched filter $s=h$,
forms exactly this sum, but it is decisive for the
BL time-encoding pipeline of Section~\ref{sec:SIrec}, which retains
only the baseband copy and therefore calls for the mixer of
Section~\ref{subsec:robustness}. For $N$ generators, stability requires
\begin{equation}
 \alpha_0\,\mathbf{I}
 \preceq \mathbf{M}_{HH}(e^{j\omega T})
 \preceq \beta_0\,\mathbf{I},
 \label{eq:riesz_multi}
\end{equation}
where $\alpha_0, \beta_0$ denote the smallest and largest eigenvalues of
$\mathbf{M}_{HH}(e^{j\omega T})$ over $|\omega|\leq\pi/T$, respectively.

\subsection{Sampling Filters and Mixers in SI Sampling}
\label{subsec:mixer_background}

Since a mixer front end is used later, we introduce it here in the classical SI
setting, alongside the prefilter, and state when each is used. A mixer replaces
the analog filter $s(t)$ by multiplication with a $T$-periodic function $p(t)=\sum_{l\in\mathbb{Z}}c_l\,e^{j2\pi lt/T}$, whose Fourier
coefficients are $c_l=\frac{1}{T}\int_{-T/2}^{T/2}p(t)e^{-j2\pi lt/T}dt$. For
$x(t)$ as in \eqref{eq:si}, modulation gives
\begin{equation}
 \mathcal{F}\{p\,x\}(\omega)=A(e^{j\omega T})\!\!\sum_{l\in\mathbb{Z}}\!
 c_l\,H\!\left(\omega-\tfrac{2\pi l}{T}\right)
 \triangleq A(e^{j\omega T})\,R(\omega),
 \label{eq:R_background}
\end{equation}
where the $T$-periodicity of $A(e^{j\omega T})$ has been used. Comparing
$R(\omega)$ with the prefilter response $R_{SH}(e^{j\omega T})$ of
\eqref{eq:rsh}, both are weighted sums of the spectral copies
$H(\omega-2\pi l/T)$: the prefilter weights them by
$\overline{S(\omega-2\pi l/T)}$ through the sampler, and the mixer weights them
by the constants $c_l$ in the analog domain. Each is inverted by a digital
correction filter, $1/R_{SH}(e^{j\omega T})$ or $1/R(\omega)$, and each requires
the same Riesz-type condition, namely that the folded response be bounded away
from zero and infinity on the band. A mixer does not weaken that requirement; it
changes which folded response one gets to design.

What separates the two devices is that summing the copies, the folding, is what
makes recovery possible at the rate $1/T$ from the band $[-\pi/T,\pi/T]$ alone,
whereas weighting them shapes the conditioning of the correction filter. A
prefilter weights within each copy but does not fold; a mixer folds but weights
each copy by a constant. The prefilter is therefore the standard choice in
classical SI sampling, where the sampler already folds all copies by itself, and
a mixer is used when the stage after the front end accepts only
a narrow band, so that the folding must take place in the analog domain before
that stage: the low-rate converter of limited analog bandwidth in the
MWC~\cite{mishali2010theory}, the modulo operator in~\cite{kvich2024modulo}, and here the BL input on which the IF-TEM recovery operates. A designed mixer avoids a filter with a prescribed wideband response~\cite{gedalyahu2011multichannel}, but the matched architecture of Corollary~\ref{cor:riesz}, presented in Section~\ref{subsec:robustness}, does not: it places the wideband filter $h(-t)$ ahead of the mixer. In hardware, each channel thus adds to the IF-TEM an LPF, a periodic waveform generator and a multiplier when a mixer is used, and, in Corollary~\ref{cor:riesz}, a filter whose passband covers the spectral copies selected by the mixer. When both
operations are needed the two are cascaded. Let the channel apply an LTI
prefilter $g(t)$, followed by the mixer $p(t)$. Since $A(e^{j\omega T})$ is
$2\pi/T$-periodic, the computation leading to \eqref{eq:R_background} gives, on
$|\omega|\leq\pi/T$,
\begin{equation}
\begin{aligned}
 \mathcal{F}\{p\cdot(g*x)\}(\omega) &= A(e^{j\omega T})\,R_{GP}(\omega),\\
 R_{GP}(\omega) &= \sum_{k\in\mathbb{Z}} c_k\,
 G\!\left(\omega-\tfrac{2\pi k}{T}\right)
 H\!\left(\omega-\tfrac{2\pi k}{T}\right).
\end{aligned}
 \label{eq:R_GP}
\end{equation}
Equation \eqref{eq:R_GP} coincides term by term with the classical
uniform-sampling response \eqref{eq:rsh} once $c_k G$ is identified with
$\overline{S}/T$: the prefilter supplies the frequency-dependent weighting
\emph{within} each spectral copy, and the mixer supplies the folding
\emph{across} copies. The classical sampler is the formal choice
$p(t)=T\sum_n\delta(t-nT)$, whose Fourier coefficients are $c_k\equiv1$, for
which \eqref{eq:R_GP} equals $T\,R_{GH}(e^{j\omega T})$ up to the conjugation
convention of \eqref{eq:rsh}. A physical mixer must use finitely many harmonics,
so that \eqref{eq:R_GP} is the classical response with the aliasing sum
truncated to those harmonics, and $R(\omega)$ of \eqref{eq:R_background} is the
further special case $G\equiv1$. In particular, a real mixer
followed by an LPF of cutoff $\pi/T$ and a uniform sampler at rate $1/T$ produces,
for every input, the samples of the classical scheme \eqref{eq:rsh} with a
sampling filter whose response is constant on each interval of width $2\pi/T$,
equal to $c_k$ on the interval centered at $2\pi k/T$. Two choices of $G$ are used in
what follows. The matched choice $G=\overline{H}$, with $c_k=1$ on the harmonics
of the mixer, turns \eqref{eq:R_GP} into the sum
$\sum_k|H(\omega-2\pi k/T)|^2$ of \eqref{eq:riesz_single} truncated to those
harmonics, and recovery is stable when this truncated sum is bounded away from zero
(Corollary~\ref{cor:riesz}). The choice $G\equiv1$, for which $R_{GP}$ reduces
to $R(\omega)$ of \eqref{eq:R_background}, avoids the analog prefilter and is
the one used in Theorems~\ref{thm:robust} and~\ref{thm:multi} and in the
designed-mixer experiments of Section~\ref{subsec:sim_mixer}; the general case amounts to replacing $R$ by $R_{GP}$.

\subsection{Problem Formulation}
\label{subsec:problem}

Let $x(t)\in\mathcal{V}$ be a $c$-bounded SI signal with $N$ generators as in
\eqref{eq:si_multi}, with the single-generator space \eqref{eq:si},
\begin{equation}
 x(t) = \sum_{n \in \mathbb{Z}} a[n]\, h(t - nT),
 \label{eq:si_signal}
\end{equation}
as an important special case. The coefficients are unknown, while the period
$T>0$ and the generators are known and satisfy the Riesz condition
\eqref{eq:riesz_multi}, which for $N=1$ is \eqref{eq:riesz_single}.

The signal is acquired by one or several IF-TEM channels. Each channel may
include an analog preprocessing stage in front of its encoder,
built from the LTI filters and periodic mixers of
Section~\ref{subsec:mixer_background}, and each IF-TEM has parameters
$(b,\kappa,\delta)$ with $b>c$. The digital side may quantize
the firing times before recovery.

Our goal is to determine the SI coefficients, and hence $x(t)$, from the firing
times of these channels and the known hardware parameters. Three requirements
guide the design. The analog front end and the digital reconstruction must be
designed jointly, since the front end decides what survives to be measured and
the reconstruction decides how to undo it. The price of that design should be
visible directly in $(b,\kappa,\delta)$, and each channel should fire close to
$T^{-1}$, so that with one channel per generator the total event rate approaches
the coefficient rate $N/T$. Finally, since no real measurement is exact, the design
must specify how noise, early termination of the recovery iteration, and
quantization of the firing times translate into coefficient error.

Two settings serve as references throughout, and every new result is measured
against them. Uniform sampling of SI signals (Section~\ref{subsec:si}) fixes
what the digital reconstruction should look like once the measurements are
available, and IF-TEM sampling of BL signals (Section~\ref{subsec:rec_IF}) fixes
what the time-encoding side looks like when the generator is a sinc, so that our
guarantees should reduce to that case exactly. Wherever a new result departs
from one of the two references, we state the departure explicitly.

\section{Time-Based Sampling and Reconstruction of SI Signals}
\label{sec:SIrec}

We begin with the simplest architecture, in which the SI signal is filtered to
the band $[-\pi/T,\pi/T]$ and handed to the IF-TEM. This costs nothing in
information whenever the generator is well behaved on the band
(Section~\ref{subsec:lpf}). Only once this single-generator, single-filter
architecture is established do we consider the generators for which it fails,
and only then is the mixer introduced as the remedy, which,
preceded by a matched prefilter, delivers stable recovery whenever the Riesz sum of the generator truncated to the mixer harmonics is bounded away from zero (Section~\ref{subsec:robustness}). The same logic is repeated one level
up for several generators sharing the channel, where the remedy is again a
mixer, now one per channel, for a reason specific to that setting
(Section~\ref{subsec:multi}). We close by comparing our results point by point
with the treatment of Gontier and Vetterli~\cite{gontier2014sampling},
the closest and most developed time-encoding treatment of SI
signals (Section~\ref{subsec:gv_relation}). The development parallels the one
carried out for modulo sampling of SI signals~\cite{kvich2024modulo}, with the
IF-TEM in the role of the modulo encoder: what changes between the two settings
is the BL recovery subroutine inside the pipeline, rather than the
architecture built around it.

\subsection{Acquisition via Low-Pass Filtering}
\label{subsec:lpf}

The recovery of Section~\ref{subsec:rec_IF} operates on BL inputs, so the analog
stage must confine the signal to a finite band before it reaches the encoder.
This is the reduction announced in Section~\ref{sec:intro},
carried out by the same stage, and for the same reason, that precedes the modulo
operator in~\cite{kvich2024modulo}. The BL recovery is interchangeable: any
recovery method for time-encoded BL signals may be placed there, and its
guarantees are inherited and expressed in $(b,\kappa,\delta)$.

The band is $[-\pi/T,\pi/T]$, the widest whose Nyquist rate equals the
degree-of-freedom rate $T^{-1}$. With a wider cutoff $\Omega'>\pi/T$, the sufficient condition \eqref{eq:density} of the BL recovery would ask the encoder to fire above $\Omega'/\pi>1/T$, so that it
captures more of the generator at a proportionally higher rate; this trade is
exercised in~\cite{kvich2024modulo}, where a cutoff of $5\pi/T$ is used for an
ECG generator. Band-limiting is thus a consequence of the reduction, and the
choice of band a consequence of matching the Nyquist rate of the encoder to the
degrees of freedom, so that the rate is set by the degrees of
freedom and not by the bandwidth of the generator. Its price is not rate but folding: unlike the
clocked sampler of \eqref{eq:rsh}, the prefilter discards the copies of $H$
outside the band instead of summing them. When those copies are not needed, an
LTI prefilter is all that is required and the pipeline reproduces classical SI
sampling exactly; when they are needed, the folding is restored by the
$T$-periodic mixer of Section~\ref{subsec:mixer_background}, the analog
counterpart of the Dirac comb, without changing the firing rate
(Section~\ref{subsec:robustness}). We therefore filter to exactly this band and
begin with the single-generator case, $N=1$.

The low-pass observation $y(t)=\mathrm{LPF}_{\pi/T}\{x(t)\}$ uniquely determines the SI coefficients whenever $H(\omega)\neq0$ almost everywhere on the retained band. Since $X(\omega) =
H(\omega)A(e^{j\omega T})$ for all $\omega$, the LPF gives
\begin{equation}
 Y(\omega) = H(\omega)\,A(e^{j\omega T}),
 \qquad |\omega| \leq \frac{\pi}{T}.
 \label{eq:Y_lpf}
\end{equation}
Under this condition, the coefficients $a[n]$ are recovered by the frequency-domain inversion
$A(e^{j\omega T}) = Y(\omega)/H(\omega)$, applied to the uniform samples
$y[nT_s]$ with $T_s=T$, whose DTFT is $Y(\omega)/T$ on the band.

In the presence of additive noise $e[nT]$, the estimated coefficients satisfy
\begin{equation}
 \hat{A}(e^{j\omega T}) = \frac{Y(\omega)+T\,E(e^{j\omega T})}{H(\omega)},
 \label{eq:A_noise}
\end{equation}
and by Parseval's identity,
\begin{equation}
 \|\hat{a}-a\|_{\ell^2}^2
 = \frac{T}{2\pi}\int_{-\pi/T}^{\pi/T}
 \frac{T^2|E(e^{j\omega T})|^2}{|H(\omega)|^2}\,d\omega.
 \label{eq:coeff_noise}
\end{equation}
Stable recovery therefore requires $H(\omega)$ to be bounded away from zero and
infinity on $|\omega|\leq\pi/T$. We call these the \emph{baseband} bounds of the
generator and write $T\alpha_{\mathrm{b}}\leq|H(\omega)|^2\leq
T\beta_{\mathrm{b}}$ a.e.\ on the band, and more generally
$T\alpha_{\mathrm{b}}\leq|S(\omega)H(\omega)|^2\leq T\beta_{\mathrm{b}}$ for a
prefilter $s(t)$ as below, the ideal LPF being the case $S(\omega)\equiv1$.
These are not the Riesz bounds \eqref{eq:riesz_single}, which constrain the
\emph{aliased} sum $R_{HH}(e^{j\omega T})$ over all spectral copies. The
baseband requirement is strictly stronger, and the two coincide,
$\alpha_{\mathrm{b}}=\alpha_0$ and $\beta_{\mathrm{b}}=\beta_0$, exactly when
$h(t)$ is baseband-confined. A perfectly valid Riesz generator may therefore
fail it, its Riesz sum being kept positive by the copies that the prefilter
discards; closing that gap is the purpose of Section~\ref{subsec:robustness},
and the gap is created by the band-limiting, not by a deficiency of the Riesz
condition.

The three ways in which a generator can meet or fail this requirement have
different consequences. When $|H(\omega)|$ is bounded away from zero, recovery
is stable and $\alpha_{\mathrm{b}}$ measures how well conditioned it is. When
$H(\omega)$ vanishes at isolated frequencies, the zero set has measure zero, the
inversion $A(e^{j\omega T})=Y(\omega)/H(\omega)$ is still defined almost
everywhere and the coefficients are still uniquely determined, but
$\alpha_{\mathrm{b}}=0$, the correction filter is unbounded and the right-hand
side of \eqref{eq:coeff_noise} need not be finite. When $H(\omega)$ vanishes on
a subset of positive measure, as for a band-pass generator, $Y(\omega)$ carries
no information about $A(e^{j\omega T})$ there. Theorem~\ref{thm:main} below
covers the first two cases and Proposition~\ref{prop:stability} the first, while
the mixer of Section~\ref{subsec:robustness} brings the second and the third
back into the framework.

Within the BL reduction, restricting to an ideal LPF is essentially without loss of generality within the
LTI class. Any prefilter $s(t)$ supported on $[-\pi/T,\pi/T]$ and non-zero a.e.\
there yields $Y(\omega)=S(\omega)H(\omega)A(e^{j\omega T})$ on the band, so it
differs from the ideal LPF only by the known invertible digital factor
$1/S(\omega)$, which is absorbed into the correction filter
$1/(S(\omega)H(\omega))$; for $S(\omega)\equiv1$ this is the classical SI
correction $1/H(\omega)$. Note that the pipeline recovers the waveform $y(t)$,
hence $Y(\omega)$ on the band, rather than the inner products $\langle
s(t-nT),x\rangle$ of uniform sampling, so the relevant response is
$S(\omega)H(\omega)$ and not the aliasing sum \eqref{eq:rsh}; the two coincide,
up to conjugation and the factor $T$, only because $S$ is supported in the band
and a single term of \eqref{eq:rsh} survives.

What no LTI prefilter followed by the LPF can do is restore the band where
$H(\omega)$ vanishes, since there $Y(\omega)=0$ regardless of $s(t)$. In uniform
SI sampling the sampler aliases the entire spectrum into \eqref{eq:rsh}, so a
suitable prefilter keeps the aliasing sum away from zero even when individual
terms vanish; the Riesz condition guarantees that such a
prefilter exists, the matched choice $s=h$ giving $R_{SH}(e^{j\omega
T})=R_{HH}(e^{j\omega T})\geq\alpha_0>0$. The BL pipeline has no such
fallback inside the LTI class, and restoring the missing aliasing requires a
linear \emph{periodically time-varying} front end, namely the analog mixer of
Section~\ref{subsec:robustness}.

\subsection{Recovery System}

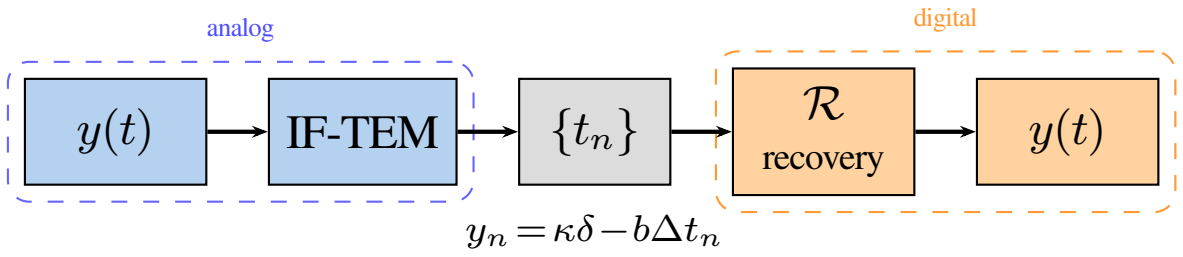
\begin{figure}[t!]
\centering
\begin{tikzpicture}[
 node distance = 1.5mm and 4mm,
 box/.style = {draw, minimum width=12mm, minimum height=7mm,
 align=center, font=\small},
 sbox/.style = {draw, minimum width=10mm, minimum height=7mm,
 align=center, font=\small},
 arr/.style = {-{Stealth[length=3pt]}, thick}
]
\node[box, fill=analogblue] (yt) {$y(t)$};
\node[box, fill=analogblue, right=of yt] (tem) {IF-TEM};
\node[sbox, fill=boundarycolor, right=of tem] (tn) {$\{t_n\}$};
\node[box, fill=digitalorg, right=of tn] (rec) {$\mathcal{R}$\\\scriptsize recovery};
\node[box, fill=digitalorg, right=of rec] (yout){$y(t)$};
\draw[arr] (yt) -- (tem);
\draw[arr] (tem) -- (tn);
\draw[arr] (tn) -- (rec);
\draw[arr] (rec) -- (yout);
\node[font=\scriptsize, below=2pt of tn] {$y_n = \kappa\delta - b\,\Delta t_n$};
\begin{scope}[on background layer]
 \node[draw=blue!60, dashed, rounded corners, fit=(yt)(tem),
 inner sep=3pt, label={[font=\tiny,blue!70]above:analog}] {};
 \node[draw=orange!80, dashed, rounded corners, fit=(rec)(yout),
 inner sep=3pt, label={[font=\tiny,orange!80]above:digital}] {};
\end{scope}
\end{tikzpicture}
\caption{IF-TEM recovery system for BL signals. Blue: analog domain.
Orange: digital domain. The inter-spike intervals
$y_n = \kappa\delta - b(t_{n+1}-t_n)$ are computed from the spike times and
used with the operator $\mathcal{R}$ to recover $y(t)$ perfectly, provided
\eqref{eq:density_condition} holds.}
\label{fig:bl_tem}
\end{figure}

Fig.~\ref{fig:bl_tem} shows the IF-TEM encoding and recovery system for a BL
signal $y(t)$, which forms the inner loop of the proposed framework. Blue
blocks indicate analog processing; orange blocks indicate digital processing.
The dashed boundary marks the analog-to-digital interface.

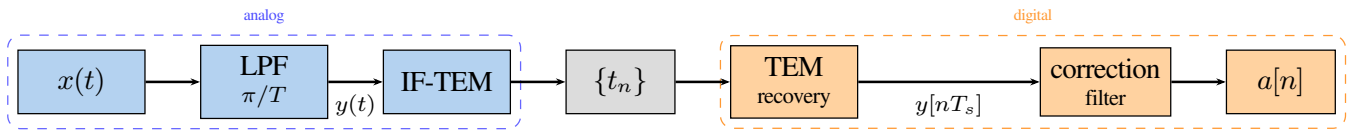
\begin{figure}[t!]
\centering
\resizebox{\linewidth}{!}{%
\begin{tikzpicture}[
 node distance = 1.5mm and 6mm,
 box/.style = {draw, minimum width=14mm, minimum height=7mm,
 align=center, font=\small},
 sbox/.style = {draw, minimum width=12mm, minimum height=7mm,
 align=center, font=\small},
 arr/.style = {-{Stealth[length=3pt]}, thick}
]
\node[box, fill=analogblue] (xt) {$x(t)$};
\node[box, fill=analogblue, right=of xt] (lpf) {LPF\\[-1pt]\scriptsize$\pi/T$};
\node[box, fill=analogblue, right=of lpf] (tem) {IF-TEM};
\node[sbox, fill=boundarycolor, right=of tem](tn) {$\{t_n\}$};
\node[box, fill=digitalorg, right=of tn] (rec) {TEM\\[-1pt]\scriptsize recovery};
\node[box, fill=digitalorg, right=20mm of rec](corr){correction\\[-1pt]\scriptsize filter};
\node[sbox, fill=digitalorg, right=of corr] (an) {$a[n]$};
\draw[arr] (xt) -- (lpf);
\draw[arr] (lpf) -- node[below,font=\scriptsize]{$y(t)$} (tem);
\draw[arr] (tem) -- (tn);
\draw[arr] (tn) -- (rec);
\draw[arr] (rec) -- node[below,font=\scriptsize]{$y[nT_s]$} (corr);
\draw[arr] (corr)-- (an);
\begin{scope}[on background layer]
 \node[draw=blue!60, dashed, rounded corners, fit=(xt)(lpf)(tem),
 inner sep=3pt, label={[font=\tiny,blue!70]above:analog}] {};
 \node[draw=orange!80, dashed, rounded corners, fit=(rec)(corr)(an),
 inner sep=3pt, label={[font=\tiny,orange!80]above:digital}] {};
\end{scope}
\end{tikzpicture}}
\caption{Block diagram for IF-TEM sampling and recovery of SI signals.
Blue: analog domain. Orange: digital domain. The SI signal $x(t)$ is
low-pass filtered to $y(t)$, encoded by the IF-TEM into firing times $\{t_n\}$,
recovered to the samples $y[nT_s]$, and passed through a correction filter to
extract the coefficients $a[n]$. See Theorem~\ref{thm:main}.}
\label{fig:si_tem}
\end{figure}

Fig.~\ref{fig:si_tem} shows the full SI recovery pipeline: $x(t)$ is
low-pass filtered to $y(t)$, encoded by the IF-TEM into $\{t_n\}$, recovered
via $\mathcal{R}$, and the SI coefficients $a[n]$ are extracted by a digital
correction filter.

\subsection{Recovery and Uniqueness}
\label{subsec:recovery}

\begin{theorem}
\label{thm:main}
Let $\mathcal{V}$ be an SI space as in \eqref{eq:si} and let
$x(t)\in\mathcal{V}$. An analog prefilter $s(t)$ whose frequency response
$S(\omega)$ is supported on $|\omega|\leq\pi/T$ is applied, yielding
$y(t)=(s*x)(t)$, and the encoder input $y(t)$ is assumed to be $c$-bounded.
The signal $y(t)$ is encoded by an IF-TEM with parameters $(b,\kappa,\delta)$,
$b>c$, producing firing times $\{t_n\}$. The ideal LPF of cutoff $\pi/T$ is
the special case $S(\omega)\equiv1$.

The signal $x(t)$ can be perfectly recovered from $\{t_n\}$, provided that
$S(\omega)H(\omega)\neq 0$ almost everywhere on $|\omega|\leq\pi/T$, and that
\begin{equation}
 \frac{\kappa\delta}{b-c} < T.
 \label{eq:density_condition}
\end{equation}
Moreover, \eqref{eq:density_condition} holds if and only if the guaranteed
firing rate $(b-c)/(\kappa\delta)$ in \eqref{eq:firing_rate_limits} exceeds the
Nyquist rate $T^{-1}$ of $y(t)$.
\end{theorem}

The boundedness hypothesis is placed on the encoder input $y(t)$ rather than on
$x(t)$, since $y(t)$ drives the integrator and the inter-firing bounds
\eqref{eq:firing_rate_limits} require $|y(t)|\leq c<b$. Note that
$\|y\|_\infty\leq\|x\|_\infty$ does not hold in general, the ideal LPF not being
non-expansive in the sup-norm. For a prefilter with $s\in L^1$ one has
$\|y\|_\infty\leq\|s\|_{1}\|x\|_\infty$, which is how the hypothesis is
certified in practice, when the implemented prefilter is bounded-input bounded-output (BIBO) stable, that is, its impulse response is absolutely integrable. The ideal brick-wall LPF is an idealization, not an exactly realizable causal filter; practical leakage and transition bands are outside the exact BL model used here. This is the same hypothesis
assumed by the BL benchmark of Section~\ref{subsec:rec_IF}, so nothing is added
here beyond the SI setting.

\begin{proof}
See Appendix~\ref{app:proofs} of the supplementary material.
\end{proof}

Theorem~\ref{thm:main} should be read against the two references of
Section~\ref{subsec:problem}. Relative to uniform SI sampling, the
identifiability condition, $H(\omega)\neq 0$ a.e.\ on the band, and the digital
correction are identical; the only difference is how the uniform samples are
obtained, namely from the firing times through the iterative scheme
\eqref{eq:iteration} instead of from a clocked sampler. Relative to IF-TEM
sampling of BL signals, condition \eqref{eq:density_condition} is exactly the
classical density condition \eqref{eq:density} with $\Omega=\pi/T$, and for
$h(t)=\sinc(\pi t/T)$ the theorem reduces to the recovery result
of~\cite{lazar2004perfect}. The condition is also sharp as a hardware
criterion: since the upper bound in \eqref{eq:interval_bounds} is approached
within the class of $c$-bounded BL inputs, \eqref{eq:density_condition} is the
weakest condition on $(b,\kappa,\delta)$ that guarantees firing above the
Nyquist rate of $y(t)$ for every signal in that class.

The assumption $H(\omega)\neq0$ a.e.\ on the band holds for many generators used
in practice and tolerates isolated in-band zeros. What the theorem does not
deliver on its own is stability: with an isolated in-band zero the coefficients
are identified while $1/H(\omega)$ is unbounded, so that
Proposition~\ref{prop:stability} below has no content, and with $|H(\omega)|$
merely small recovery is stable but poorly conditioned. All three situations are
addressed by the mixer of Section~\ref{subsec:robustness}.

\subsection{Stability Analysis}
\label{subsec:stability}

Theorem~\ref{thm:main} guarantees perfect recovery in the limit $l\to\infty$.
In practice the iteration \eqref{eq:iteration} is stopped after $l$ steps,
giving an approximation $\hat{y}(t)$ with error bounded by
\eqref{eq:convergence}. The following result quantifies how this
finite-iteration error propagates through the SI inversion.

\begin{proposition}
\label{prop:stability}
Under the conditions of Theorem~\ref{thm:main}, assume the baseband bound
$|S(\omega)H(\omega)|^2 \geq T\alpha_{\mathrm{b}} > 0$ a.e.\ on
$|\omega|\leq\pi/T$. Let $\hat{y}(t)$ satisfy
$\|\hat{y}-y\|_{L^2}\leq\varepsilon$. Then the recovered coefficients
$\{\hat{a}[n]\}$ via \eqref{eq:A_recovery} satisfy
\begin{equation}
 \|\hat{a} - a\|_{\ell^2}
 \leq
 \frac{1}{\sqrt{\alpha_{\mathrm{b}}}}\,\varepsilon.
 \label{eq:si_error}
\end{equation}
In particular, after $l$ iterations, $\varepsilon = r^{l+1}\|y\|_{L^2}$ with
$r = \kappa\delta\Omega/((b-c)\pi)$, giving
\begin{equation}
 \|\hat{a} - a\|_{\ell^2}
 \leq
 \frac{1}{\sqrt{\alpha_{\mathrm{b}}}}\,r^{l+1}\|y\|_{L^2}.
 \label{eq:si_error_iter}
\end{equation}
\end{proposition}

Here $\alpha_{\mathrm{b}}$ is the baseband bound of Section~\ref{subsec:lpf} and
not the Riesz bound $\alpha_0$; for a wideband generator
\eqref{eq:riesz_single} does \emph{not} imply $\alpha_{\mathrm{b}}>0$, and the
proposition then does not apply however well conditioned the SI space itself is.

\begin{proof}
See Appendix~\ref{app:proofs} of the supplementary material.
\end{proof}

The bound \eqref{eq:si_error_iter} exposes two independent factors. The IF-TEM
factor $r=\kappa\delta/((b-c)T)<1$ depends entirely on hardware: decreasing
$\kappa\delta$ or increasing the bias margin $(b-c)$ improves convergence,
exactly as in the BL setting \eqref{eq:convergence}. The generator factor
$1/\sqrt{\alpha_{\mathrm{b}}}$ is the new ingredient relative to that setting
and quantifies the price of the SI inversion. The generator enters only through
$\alpha_{\mathrm{b}}$, so two generators with the same baseband lower bound
incur the same worst-case amplification whatever their shape; the out-of-band
behavior that the Riesz bound $\alpha_0$ accounts for is invisible here because
the prefilter has already discarded it.

The comparison with the BL benchmark is exact. For $h(t)=\sinc(\pi t/T)$ the
generator is baseband-confined, $\alpha_{\mathrm{b}}=\alpha_0=T$, and
\eqref{eq:si_error} reads $\|\hat a-a\|_{\ell^2}\leq\varepsilon/\sqrt{T}$. Since
then $a[n]=y(nT)$ and $\|\hat y-y\|_{L^2}^2=T\|\hat a-a\|_{\ell^2}^2$ by
Parseval's theorem for BL signals sampled at the Nyquist rate, the bound holds
with equality when $\|\hat y-y\|_{L^2}=\varepsilon$.
Proposition~\ref{prop:stability} is thus tight in the BL case and states that a
general generator degrades the BL baseline by exactly
$\sqrt{T/\alpha_{\mathrm{b}}}$. The proof splits accordingly: the geometric
decay of the finite-iteration error is inherited from the BL
theory~\cite{lazar2004perfect}, while the new element is the propagation of that
error through the correction filter, controlled by $\alpha_{\mathrm{b}}$.

As Section~\ref{subsec:lpf} showed, what no prefilter
supported in the band can supply is the folding itself, and that is the gap
addressed next by the mixer.

\subsection{Restoring the Folding with a Mixer}
\label{subsec:robustness}

The three failure modes of Section~\ref{subsec:lpf} share one feature: the
generator may still form a Riesz basis for the SI space, its Riesz sum
\eqref{eq:riesz_single} drawing on spectral copies that the low-pass filter has
already discarded. Stated in the form used for modulo sampling of SI
signals~\cite{kvich2024modulo}, a generator can form a Riesz basis while its
low-pass-filtered version does not. A concrete instance is the scaled
first-order B-spline, a valid SI Riesz generator whose baseband response is a
$\mathrm{sinc}^2$ with a root inside $[-\pi/T,\pi/T]$: the LPF keeps only that
copy, so $\mathrm{LPF}\{h\}$ fails the Riesz condition and $1/H(\omega)$
diverges, whereas the full generator, and the classical uniform sampler that
aliases all its copies, do not.

\begin{figure}[t!]
\centering
\resizebox{\linewidth}{!}{%
\begin{tikzpicture}[
 node distance = 1.5mm and 4mm,
 box/.style = {draw, minimum width=11mm, minimum height=6mm,
 align=center, font=\scriptsize},
 sbox/.style = {draw, minimum width=9mm, minimum height=6mm,
 align=center, font=\scriptsize},
 arr/.style = {-{Stealth[length=2.5pt]}, thick}
]
\node[box, fill=analogblue] (xt) {$x(t)$};
\node[box, fill=analogblue, right=of xt] (mix) {mixer\\[-1pt]\tiny$p(t)$};
\node[box, fill=analogblue, right=14mm of mix] (lpf) {LPF\\[-1pt]\tiny$\pi/T$};
\node[box, fill=analogblue, right=14mm of lpf] (tem) {IF-TEM};
\node[sbox, fill=boundarycolor, right=of tem] (tn) {$\{t_n\}$};
\node[box, fill=digitalorg, right=of tn] (rec) {TEM\\[-1pt]\tiny recovery};
\node[box, fill=digitalorg, right=14mm of rec] (corr) {correction\\[-1pt]\tiny filter};
\node[sbox, fill=digitalorg, right=of corr] (an) {$a[n]$};
\draw[arr] (xt) -- (mix);
\draw[arr] (mix) -- node[below,font=\tiny]{$p(t)x(t)$} (lpf);
\draw[arr] (lpf) -- node[below,font=\tiny]{$y(t)$} (tem);
\draw[arr] (tem) -- (tn);
\draw[arr] (tn) -- (rec);
\draw[arr] (rec) -- node[below,font=\tiny]{$y[nT_s]$} (corr);
\draw[arr] (corr) -- (an);
\begin{scope}[on background layer]
 \node[draw=blue!60, dashed, rounded corners, fit=(xt)(mix)(lpf)(tem),
 inner sep=3pt, label={[font=\tiny,blue!70]above:analog}] {};
 \node[draw=orange!80, dashed, rounded corners, fit=(rec)(corr)(an),
 inner sep=3pt, label={[font=\tiny,orange!80]above:digital}] {};
\end{scope}
\end{tikzpicture}}
\caption{Block diagram for IF-TEM sampling and recovery of SI signals with
a mixer that restores the spectral folding. Blue: analog
domain. Orange: digital domain.
A mixer with $T$-periodic signal $p(t)$ is inserted before the LPF,
changing the digital correction on the Nyquist samples from $T/H(\omega)$ to $T/R(\omega)$
as defined in \eqref{eq:R_omega}. See Theorem~\ref{thm:robust}.}
\label{fig:si_tem_mixer}
\end{figure}
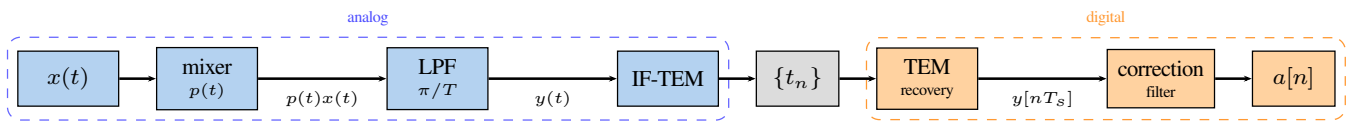

The mixer addresses all three modes by a single mechanism, restoring in the
analog domain the folding that a clocked sampler performs at no cost and that
band-limiting has removed. We therefore apply a $T$-periodic analog mixer $p(t)$
to $x(t)$ before the LPF and the encoding, as in Fig.~\ref{fig:si_tem_mixer} and
as in the mixer-based stability enhancement of~\cite{kvich2024modulo}. The
device is the one introduced in Section~\ref{subsec:mixer_background}: let
$p(t)=\sum_{l\in\mathbb{Z}}c_l\,e^{j2\pi lt/T}$ be real valued, with Fourier
coefficients $c_l=\frac{1}{T}\int_{-T/2}^{T/2}p(t)\,e^{-j2\pi lt/T}\,dt$, and
recall from \eqref{eq:R_background}
\begin{equation}
 R(\omega) = \sum_{l\in\mathbb{Z}} c_l\,H\!\left(\omega-\frac{2\pi l}{T}\right).
 \label{eq:R_omega}
\end{equation}
The condition required below, $R(\omega)\neq0$ a.e.\ on the band, is the
Riesz-type condition of Section~\ref{subsec:mixer_background} with $R(\omega)$
in the role of $R_{SH}(e^{j\omega T})$, and the Riesz condition
on $h$ is what makes it attainable. By \eqref{eq:riesz_single},
$\sum_k|H(\omega-2\pi k/T)|^2\geq T\alpha_0>0$ for a.e.\ $\omega$, so at least
one spectral copy is non-zero at a.e.\ $\omega$, even where the baseband copy
vanishes, and the information discarded by the LPF is present in some other
copy. Since the $c_l$ are constants while the copies may vanish at different
frequencies, the condition to impose on a mixer with finite harmonic support
$\mathcal{K}$ is the finite-harmonic form of the same requirement, as in the
corresponding step for modulo sampling of SI
spaces~\cite[Sec.~3.3]{kvich2024modulo}. The folded response then factors as
\begin{equation}
 R(\omega) = \mathbf{c}^{\mathsf{T}}\mathbf{h}(\omega), \quad
 \mathbf{c} = [c_k]_{k\in\mathcal{K}}, \quad
 \mathbf{h}(\omega) =
 \left[H\!\left(\omega-\frac{2\pi k}{T}\right)\right]_{k\in\mathcal{K}},
 \label{eq:R_factor}
\end{equation}
with the constant mixer coefficients on one side and the
generator, evaluated on the harmonic grid, on the other. If for a.e.\
$\omega\in[-\pi/T,\pi/T]$ some $k\in\mathcal{K}$ has $H(\omega-2\pi k/T)\neq0$,
an appropriate choice of $\{c_k\}$ yields $R(\omega)\neq0$ a.e.; this is the
single-channel case of the rank condition of~\cite{kvich2024modulo}, and the
mixer is selected for a given generator and verified by evaluating $R(\omega)$
on a fine grid over the band. A positive minimum on a finite grid is a numerical diagnostic, not a certificate of stability. A uniform lower bound requires an analytic argument or a validated bound on the variation between grid points. For the scaled first-order B-spline of
Section~\ref{subsec:sim_mixer} the five harmonics $\mathcal{K}=\{-2,\ldots,2\}$
already suffice, and Corollary~\ref{cor:riesz} below removes the design step
altogether by placing a matched prefilter ahead of a mixer with equal
coefficients. The mixer thus does not weaken the Riesz requirement, it makes
the pipeline able to use it, spreading the energy of $H(\omega)$ across
frequency so that $1/R(\omega)$ can be well conditioned even where $H(\omega)$
has in-band zeros.

\begin{theorem}
\label{thm:robust}
Let $\mathcal{V}$ be an SI space as in \eqref{eq:si} and let
$x(t)\in\mathcal{V}$. A mixer with real $T$-periodic $p(t)$ and an LPF with
cutoff $\pi/T$ are applied, yielding
$y(t)=\mathrm{LPF}_{\pi/T}\{p(t)x(t)\}$, and the encoder input $y(t)$ is
assumed to be $c$-bounded as in Theorem~\ref{thm:main}. The signal $y(t)$ is
encoded by an IF-TEM with parameters $(b,\kappa,\delta)$, $b>c$, producing
$\{t_n\}$.

The signal $x(t)$ can be perfectly recovered from $\{t_n\}$, provided that
$R(\omega)$ as defined in \eqref{eq:R_omega} is non-zero almost everywhere on
$|\omega|\leq\pi/T$, and that condition \eqref{eq:density_condition} holds.
\end{theorem}

As in Theorem~\ref{thm:main}, the ideal LPF may be replaced by any prefilter
$s(t)$ supported on $[-\pi/T,\pi/T]$: the condition becomes
$S(\omega)R(\omega)\neq0$ a.e.\ and $S(\omega)$ is absorbed into the correction
filter. The essential element here is the mixer, which is periodically time
varying, and not the particular LTI stage that follows it.

\begin{proof}
See Appendix~\ref{app:proofs} of the supplementary material.
\end{proof}

Setting $p(t)=1$ gives $R(\omega)=H(\omega)$, so Theorem~\ref{thm:robust}
contains Theorem~\ref{thm:main} as a special case. The freedom in choosing $p(t)$ shapes $R(\omega)$ so
that it avoids the zeros of $H(\omega)$ and distributes energy more evenly
across the band, and it improves the conditioning of the folded response, which
controls noise amplification exactly as $\alpha_{\mathrm{b}}$ does in
Proposition~\ref{prop:stability} with $H$ replaced by $R$. The mixer restores identifiability when $R(\omega)\neq0$ almost everywhere, and stable inversion when $\operatorname*{ess\,inf}_{|\omega|\leq\pi/T}|R(\omega)|>0$, without changing the sufficient firing-density condition; in practice $p(t)$ is chosen from knowledge of $H(\omega)$ and the noise
level, as in~\cite{kvich2024modulo}.

Read against the two reference settings, Theorem~\ref{thm:robust} has no
counterpart in BL time encoding, where the generator is the sinc and the
response seen by the encoder is flat. Against uniform SI sampling it plays the
part of the choice of sampling filter, both shaping the response that the
digital correction inverts, with the difference that the mixer must act before
the encoder rather than at the sampler. The proof is that of
Theorem~\ref{thm:main} with $H(\omega)$ replaced by $R(\omega)$, so what is new
is the architecture and the condition it makes attainable, rather than the
argument.

Placing the matched prefilter of
Section~\ref{subsec:mixer_background} ahead of a mixer with equal coefficients
makes the folded response equal to the Riesz sum of the generator truncated to the mixer harmonics, so that no mixer needs to be designed.

\begin{cor}[Riesz generators]
\label{cor:riesz}
Let $\mathcal{V}$ be an SI space as in \eqref{eq:si} whose generator $h(t)$
satisfies the Riesz condition \eqref{eq:riesz_single}, let
$x(t)\in\mathcal{V}$, and let $\mathcal{K}\subset\mathbb{Z}$ be a finite set of
harmonics with $\mathcal{K}=-\mathcal{K}$. Define
\begin{equation}
 R_{\mathcal{K}}(\omega)=\sum_{k\in\mathcal{K}}
 \Bigl|H\!\left(\omega-\tfrac{2\pi k}{T}\right)\Bigr|^2,
 \qquad
 \alpha_{\mathcal{K}}=\operatorname*{ess\,inf}_{|\omega|\leq\pi/T}
 \frac{R_{\mathcal{K}}(\omega)}{T}.
 \label{eq:RK}
\end{equation}
The matched prefilter $g(t)=h(-t)$, the mixer
$p(t)=\sum_{k\in\mathcal{K}}e^{j2\pi kt/T}$ and an LPF with cutoff $\pi/T$ are
applied, yielding $y(t)=\mathrm{LPF}_{\pi/T}\{p\cdot(g*x)\}(t)$, which is
assumed to be $c$-bounded and is encoded by an IF-TEM with parameters
$(b,\kappa,\delta)$, $b>c$, satisfying \eqref{eq:density_condition}. If
$\alpha_{\mathcal{K}}>0$, then $x(t)$ is perfectly recovered from $\{t_n\}$
through the correction filter $1/R_{\mathcal{K}}(\omega)$, and an estimate
$\hat{y}(t)$ with $\|\hat{y}-y\|_{L^2}\leq\varepsilon$ yields coefficients
satisfying
\begin{equation}
 \|\hat{a}-a\|_{\ell^2}\leq\frac{\varepsilon}{\sqrt{T}\,\alpha_{\mathcal{K}}}.
 \label{eq:riesz_stability}
\end{equation}
If, moreover, $H(\omega)$ and $R_{HH}(e^{j\omega T})$ are continuous, then
$\alpha_{\mathcal{K}}$ increases to the optimal lower Riesz bound
$\alpha_0=\min_\omega R_{HH}(e^{j\omega T})$ as $\mathcal{K}$ grows to
$\mathbb{Z}$ through the sets $\{-m,\ldots,m\}$.
\end{cor}

\begin{proof}
See Appendix~\ref{app:proofs} of the supplementary material.
\end{proof}

Corollary~\ref{cor:riesz} reads directly against the two
references. Against uniform SI sampling, the matched prefilter and the mixer
realize the matched sampling filter $s=h$ with the aliasing sum \eqref{eq:rsh}
truncated to $\mathcal{K}$. The samples $y[nT]$ have DTFT $A(e^{j\omega T})R_{\mathcal{K}}(\omega)/T$, which for $\mathcal{K}=\mathbb{Z}$ is the DTFT $R_{HH}(e^{j\omega T})A(e^{j\omega T})$ of the classical samples $c[n]$ with $s=h$. Since $y(t)$ is sampled at its Nyquist rate, $\|\hat{y}-y\|_{L^2}=\sqrt{T}\,\|\hat{y}[nT]-y[nT]\|_{\ell^2}$, and \eqref{eq:riesz_stability} reads $\|\hat{a}-a\|_{\ell^2}\leq\|\hat{y}[nT]-y[nT]\|_{\ell^2}/\alpha_{\mathcal{K}}$. This is the uniform-sampling bound $\|\hat{a}-a\|_{\ell^2}\leq\|\hat{c}-c\|_{\ell^2}/\alpha_0$ with $\alpha_{\mathcal{K}}$ in place of $\alpha_0$. For the IF-TEM, the finite-iteration error in these samples is controlled by $r=\kappa\delta/((b-c)T)$ in \eqref{eq:convergence}. Against BL time encoding, the sinc generator needs
$\mathcal{K}=\{0\}$, for which $g$ is the ideal LPF scaled by $T$, $p(t)=1$, and
$\alpha_{\mathcal{K}}=\alpha_0=T$. The two architectures above are the cases in
which part of the construction can be dropped. When the baseband copy is bounded
away from zero, $\mathcal{K}=\{0\}$ already suffices and the matched prefilter
can be replaced by the LPF alone, which is the setting of
Theorem~\ref{thm:main} and Proposition~\ref{prop:stability}. When it is not, the
prefilter can be dropped in favor of a mixer whose coefficients are designed for
the given generator, as in Theorem~\ref{thm:robust} and in the experiments of
Section~\ref{subsec:sim_mixer}.

\subsection{Extension to Multiple Generators}
\label{subsec:multi}

We now extend the framework to SI spaces with $N$ generators as in
\eqref{eq:si_multi}, using multiple channels, each with its own analog mixer and
IF-TEM, as in the multichannel approach of~\cite{kvich2024modulo}.

In uniform multichannel SI sampling both analog options are classical: a bank of
sampling filters leads to inversion of $\mathbf{M}_{HH}$ in
\eqref{eq:riesz_multi}~\cite{eldar2015sampling}, while banks of periodic
modulators are used in the MWC~\cite{mishali2010theory} and in modulo sampling of SI
signals~\cite{kvich2024modulo}. In the time-encoding pipeline only the mixer
option is viable. If channel $i$ applied an LTI prefilter $S_i(\omega)$ followed
by the band-limiting LPF, then on the band
$Y_i(\omega)=S_i(\omega)\sum_{l=1}^{N}H_l(\omega)A_l(e^{j\omega T})$, so the
matrix relating the coefficients to the channel outputs is
$[S_i(\omega)H_l(\omega)]_{i,l}$, an outer product of rank at most one at every
$\omega$; no choice of LTI filters ahead of the LPF separates $N\geq2$ generators. A bank of \emph{distinct} mixers removes the obstruction, each mixer
folding a different weighted combination of the spectral copies into the band.
The mixers therefore play a twofold role: for $N=1$ a single mixer reshapes the
conditioning (Theorem~\ref{thm:robust}), while for $N\geq2$ the bank
additionally separates the generators.

Let the mixer in the $i$-th channel be $p_i(t)$, with Fourier coefficients
$\{c_{i,k}\}_{k\in\mathbb{Z}}$. Define the $I \times N$ matrix-valued
function
\begin{equation}
 \mathbf{M}(\omega) =
 \left[
 \sum_{k\in\mathbb{Z}} c_{i,k}\, H_l\!\left(\omega - \frac{2\pi k}{T}\right)
 \right]_{i=1,\ldots,I;\; l=1,\ldots,N},
 \label{eq:M_omega}
\end{equation}
which reduces to the scalar $R(\omega)$ of \eqref{eq:R_omega} when $N=I=1$.

\begin{theorem}
\label{thm:multi}
Let $\mathcal{V}$ be an SI space with $N$ generators as in \eqref{eq:si_multi}
and let $x(t)\in\mathcal{V}$. In each channel $i=1,\ldots,I$ with $I\geq N$, a
mixer with real $T$-periodic $p_i(t)$ and an LPF with cutoff $\pi/T$ are
applied, yielding $y_i(t) = \mathrm{LPF}_{\pi/T}\{p_i(t)x(t)\}$, and each
encoder input $y_i(t)$ is assumed to be $c$-bounded as in
Theorem~\ref{thm:main}. Each $y_i(t)$ is encoded by an IF-TEM with parameters
$(b,\kappa,\delta)$, $b>c$, producing firing times $\{t_n^{(i)}\}$, and the
density condition \eqref{eq:density_condition} holds in each channel.

The signal $x(t)$ can be perfectly recovered from $\{t_n^{(i)}\}_{i=1}^{I}$,
provided that $\mathbf{M}(\omega)$ as defined in \eqref{eq:M_omega} has rank
at least $N$ almost everywhere on $|\omega|\leq\pi/T$.
\end{theorem}

\begin{proof}
See Appendix~\ref{app:proofs} of the supplementary material.
\end{proof}

\begin{figure*}[t!]
\centering
\resizebox{\linewidth}{!}{%
\begin{tikzpicture}[
 box/.style = {draw, minimum width=18mm, minimum height=9mm,
 align=center, font=\scriptsize},
 sbox/.style = {draw, minimum width=16mm, minimum height=9mm,
 align=center, font=\scriptsize},
 arr/.style = {-{Stealth[length=3pt]}, thick}
]

\def\ytop{1.4}
\def\ybot{-1.4}
\def\ymid{0}
\def\xmix{3.2}
\def\xlpf{6.4}
\def\xtem{9.4}
\def\xspk{11.8}
\def\xrec{14.0}
\def\xcor{16.8}
\def\xout{19.4}

\node[box, fill=analogblue] at (0, \ymid) (x) {$x(t)$};

\node[box, fill=analogblue] at (\xmix,\ytop) (m1) {mixer\\[-2pt]\tiny$p_1(t)$};
\node[box, fill=analogblue] at (\xlpf,\ytop) (l1) {LPF\\[-2pt]\tiny$\pi/T$};
\node[box, fill=analogblue] at (\xtem,\ytop) (t1) {IF-TEM};
\node[sbox,fill=boundarycolor] at (\xspk,\ytop) (s1) {$\{t_n^{(1)}\}$};
\node[box, fill=digitalorg] at (\xrec,\ytop) (r1) {TEM\\[-2pt]\tiny recovery};

\node[box, fill=analogblue] at (\xmix,\ybot) (mI) {mixer\\[-2pt]\tiny$p_I(t)$};
\node[box, fill=analogblue] at (\xlpf,\ybot) (lI) {LPF\\[-2pt]\tiny$\pi/T$};
\node[box, fill=analogblue] at (\xtem,\ybot) (tI) {IF-TEM};
\node[sbox,fill=boundarycolor] at (\xspk,\ybot) (sI) {$\{t_n^{(I)}\}$};
\node[box, fill=digitalorg] at (\xrec,\ybot) (rI) {TEM\\[-2pt]\tiny recovery};

\node[font=\large] at (\xmix, \ymid) {$\vdots$};

\node[box, fill=digitalorg] at (\xcor,\ymid) (corr)
 {correction filter\\[-2pt]\tiny$\mathbf{M}^\dagger(\omega)$};
\node[sbox,fill=digitalorg] at (\xout,\ymid) (an) {$\{a_l[n]\}$};

\draw[arr] (x.east) -- ++(0.8,0) |- (m1.west);
\draw[arr] (x.east) -- ++(0.8,0) |- (mI.west);
\draw[arr] (m1.east) -- node[above,font=\tiny]{$p_1(t)x(t)$} (l1.west);
\draw[arr] (l1.east) -- node[above,font=\tiny]{$y_1(t)$} (t1.west);
\draw[arr] (t1.east) -- (s1.west);
\draw[arr] (s1.east) -- (r1.west);
\draw[arr] (mI.east) -- node[above,font=\tiny]{$p_I(t)x(t)$} (lI.west);
\draw[arr] (lI.east) -- node[above,font=\tiny]{$y_I(t)$} (tI.west);
\draw[arr] (tI.east) -- (sI.west);
\draw[arr] (sI.east) -- (rI.west);
\draw[arr] (r1.east) -- ++(0.7,0) |- (corr.north);
\draw[arr] (rI.east) -- ++(0.7,0) |- (corr.south);
\draw[arr] (corr.east) -- (an.west);

\begin{scope}[on background layer]
 \draw[draw=blue!50, dashed, rounded corners=5pt, fill=analogblue!25,
 line width=0.6pt]
 (-1.2, 2.2) rectangle (10.6, -2.2);
 \node[font=\tiny\bfseries, blue!70, anchor=north west]
 at (-1.2, 2.2) {analog};
 \draw[draw=blue!40, rounded corners=3pt, line width=0.5pt]
 (2.0, 2.05) rectangle (15.2, 0.65);
 \node[font=\tiny, blue!60, anchor=south]
 at (8.6, 2.05) {Channel~$1$};
 \draw[draw=blue!40, rounded corners=3pt, line width=0.5pt]
 (2.0, -0.65) rectangle (15.2, -2.05);
 \node[font=\tiny, blue!60, anchor=north]
 at (8.6,-2.05) {Channel~$I$};
 \draw[draw=orange!60, dashed, rounded corners=5pt, fill=digitalorg!25,
 line width=0.6pt]
 (12.8, 2.2) rectangle (20.6, -2.2);
 \node[font=\tiny\bfseries, orange!80, anchor=north east]
 at (20.6, 2.2) {digital};
\end{scope}

\end{tikzpicture}}
\caption{Block diagram for IF-TEM sampling and recovery of SI signals with
$N$ generators using $I$ channels. Blue: analog domain; orange: digital
domain. The input $x(t)$ is broadcast to all $I$ channels. Each channel
applies its own analog mixer $p_i(t)$, LPF, and IF-TEM independently.
The channel outputs are combined in the digital domain via the pseudoinverse
correction filter $\mathbf{M}^\dagger(\omega)$ to recover the SI
coefficients $\{a_l[n]\}$. See Theorem~\ref{thm:multi}.}
\label{fig:si_tem_multi}
\end{figure*}
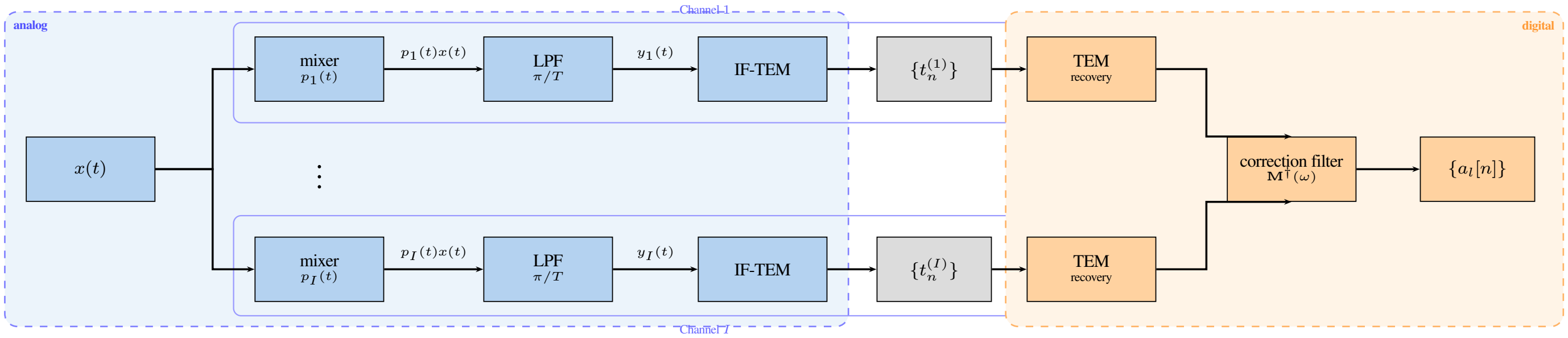

As for Theorem~\ref{thm:main}, the rank condition gives identifiability but not stability: if the smallest singular value of $\mathbf{M}(\omega)$ is not bounded away from zero on the band, the pseudoinverse in \eqref{eq:multi_recovery} is unbounded. Stability requires the uniform lower bound \eqref{eq:sigma_lower} of Proposition~\ref{prop:stability_multi}. The mixers $\{p_i(t)\}$, which must differ across channels since identical
mixers reproduce the rank-one obstruction above, are chosen so that
$\mathbf{M}(\omega)$ has rank $N$ and a favorable condition number across the
band. Only the digital correction step, the pseudoinverse of
$\mathbf{M}(\omega)$, differs from the single-generator case.
The resulting architecture resembles the MWC~\cite{mishali2010theory}, as the multichannel mixer bank
of~\cite{kvich2024modulo} does, and is adapted here to a different signal class
and to an event-based encoder in place of a low-rate uniform sampler. Each
channel fires just above $T^{-1}$, so with $I=N$ channels the total event rate
approaches $N/T$, and for $N=I=1$ with $p_1(t)=1$ Theorem~\ref{thm:multi}
reduces to Theorem~\ref{thm:main}. The block diagram is shown in
Fig.~\ref{fig:si_tem_multi}.

\begin{proposition}
\label{prop:stability_multi}
Under the conditions of Theorem~\ref{thm:multi}, assume there exists a
constant $\sigma > 0$ such that
\begin{equation}
 \frac{\sigma_{\min}(\omega)}{\sqrt{T}} \geq \sigma
 \quad \text{for almost all } |\omega| \leq \frac{\pi}{T},
 \label{eq:sigma_lower}
\end{equation}
where $\sigma_{\min}(\omega)$ is the smallest singular value of
$\mathbf{M}(\omega)$. Let $\hat{y}_i(t)$ satisfy
$\|\hat{y}_i - y_i\|_{L^2} \leq \varepsilon$ for each channel $i$. Then
the recovered coefficient vectors $\{\hat{a}_l[n]\}_{l=1}^N$ satisfy
\begin{equation}
 \sum_{l=1}^N \|\hat{a}_l - a_l\|_{\ell^2}^2
 \leq \frac{1}{\sigma^2}
 \sum_{i=1}^I \|\hat{y}_i - y_i\|_{L^2}^2
 \leq \frac{I}{\sigma^2}\,\varepsilon^2.
 \label{eq:multi_stability}
\end{equation}
\end{proposition}

\begin{proof}
See Appendix~\ref{app:proofs} of the supplementary material.
\end{proof}

For $N=I=1$ and $p_1(t)=1$, so that $\mathbf{M}(\omega)$ reduces to the scalar
$H(\omega)$, condition \eqref{eq:sigma_lower} becomes $|H(\omega)|/\sqrt{T}\geq
\sigma$ a.e., which is the baseband bound $\sigma=\sqrt{\alpha_{\mathrm{b}}}$ of
Proposition~\ref{prop:stability}, and \eqref{eq:multi_stability} recovers that
proposition. As in the single-generator case, \eqref{eq:sigma_lower} constrains
the baseband response of the mixed generators and is \emph{not} implied by the
multi-generator Riesz condition \eqref{eq:riesz_multi}, which constrains the
aliased matrix $\mathbf{M}_{HH}$; it is the matrix analogue of
$\alpha_{\mathrm{b}}>0$, and it does not follow from the rank condition of Theorem~\ref{thm:multi}; the mixers must be chosen to secure it.

Relative to uniform multichannel SI sampling, Theorem~\ref{thm:multi} and
Proposition~\ref{prop:stability_multi} are the familiar statements with
$\mathbf{M}(\omega)$ in the role of $\mathbf{M}_{HH}(e^{j\omega T})$ and its
smallest singular value in the role of the lower Riesz bound, the proofs being
the single-generator arguments applied channel-wise followed by a pseudoinverse.
Relative to BL time encoding there is no counterpart, a BL space being generated
by a single function. What differs from both is the necessity of the mixers,
which the rank-one obstruction makes an architectural constraint rather than a
design preference.

\subsection{Relation to the Framework of Gontier and Vetterli}
\label{subsec:gv_relation}

With the results in place, we compare with the treatment of TEMs on SI subspaces
by Gontier and Vetterli~\cite{gontier2014sampling}, which establishes
invertibility and iterative reconstruction for crossing-based and
integrate-and-fire encoders. Both treatments require the shifts
of the generator to form a Riesz basis. They differ in the route from firing
times to coefficients, in what sets the event density, and in what the encoder
records.

The first difference is the route.
In~\cite{gontier2014sampling} the encoder is treated as a nonuniform sampler and
the firing times are inverted directly in the SI space, through an operator
built from the generator and the firing set. Here the SI signal is mapped in the
analog domain onto the BL class and the existing BL inversion is applied, the
design pattern already used for streams of
pulses~\cite{naaman2021fri,alexandru2019reconstructing}. The convergence and
quantization analysis of the BL theory then transfers unchanged, and the BL
stage is interchangeable, so any other BL recovery can be substituted without
touching the SI layer. What the route costs is the folding, which the mixer
returns (Corollary~\ref{cor:riesz}). It also costs an analog front end that the direct route of~\cite{gontier2014sampling} does not need: an LPF in every channel, a mixer when the baseband copy does not suffice, and, in Corollary~\ref{cor:riesz}, a matched prefilter.

The second difference is what sets the event density.
By~\cite[Sec.~4]{gontier2014sampling}, an IF-TEM whose longest inter-spike
interval satisfies $\Delta t_{\max}<\tau$ is invertible, where
\begin{equation}
 \tau=\pi\inf_{\omega}\frac{G_h(\omega)}{G_{h'}(\omega)},\qquad
 G_h(\omega)=\Bigl(\sum_{k\in\mathbb{Z}}
 \bigl|H\!\left(\omega-\tfrac{2\pi k}{T}\right)\bigr|^2\Bigr)^{1/2},
 \label{eq:gv_tau}
\end{equation}
and $G_{h'}$ is defined in the same way for the derivative $h'$, which is
required to be square integrable. With $\Delta t_{\max}=\kappa\delta/(b-c)$ from
\eqref{eq:density}, the condition reads $\kappa\delta/(b-c)<\tau$. The threshold
$\tau$ equals $T$ for the sinc generator, is in general smaller for generators
with energy beyond the band, and for a band-pass generator centered at a high
frequency it is of the order of the inverse Nyquist rate of the generator. Here
the condition is $\kappa\delta/(b-c)<T$ for every generator, and the generator
enters only through the conditioning of the correction filter, in
Proposition~\ref{prop:stability}, Corollary~\ref{cor:riesz} and
Theorem~\ref{thm:quant_si}, as in uniform SI sampling. The band-limiting front
end also accepts generators of any smoothness, such as the zeroth-order
B-spline, whose derivative is not square integrable. In the experiment of Section~\ref{subsec:sim_compare}, with the same IF-TEM and about $2.4$ events per period, direct inversion is $0.7$--$1.0$~dB more accurate for a cubic spline under timing error, and our front end lowers the coefficient error of a band-pass generator by about $25$~dB.

The third difference is what the encoders record. The crossing
construction of~\cite{gontier2014sampling} reads the signal amplitude at the
instants where the input crosses a test function, whereas the IF-TEM records
inter-spike intervals through \eqref{eq:measurement_y_n}. Since
timing error is the dominant impairment in clockless hardware, we compare the two empirically: at a matched event rate and behind the same LPF, the two methods are within a few dB when the
firing times are nearly exact, the crossing encoder being marginally better,
whereas once the timing-error standard deviation exceeds roughly $10^{-3}T$ the
proposed pipeline improves the coefficient NMSE by $6$-$10$~dB for a BL
generator and by $4$-$6$~dB for a cubic-spline generator
(Section~\ref{subsec:sim_compare}, Table~\ref{tab:compare_gv}).

Two further elements have no counterpart there, namely the SI
spaces with multiple generators of Section~\ref{subsec:multi}, and an explicit
mean-squared-error bound for quantized firing times.
In~\cite[Sec.~6]{gontier2014sampling} the coefficient error of the crossing
encoder is shown to grow at most linearly with a bounded timing error, through a
constant that depends on the generator and is not given in closed form, whereas
Section~\ref{sec:quant} bounds the error in $(b,\kappa,\delta)$ and in the
conditioning of the folded response and places it side by side with amplitude
quantization at the same bit depth per measurement.

\section{Quantization Analysis}
\label{sec:quant}

In practical hardware, the firing times $\{t_n\}$ must be stored with finite
precision. We analyze the effect of uniformly quantizing the inter-spike
intervals $T_n = t_{n+1} - t_n$ and derive an explicit MSE bound for the
recovered SI coefficients.

\subsection{Setup and Quantization Step Size}
\label{subsec:quant_setup}

The signal entering the IF-TEM is $y(t) = \mathrm{LPF}_{\pi/T}\{x(t)\}$,
which is $c$-bounded and $2\Omega$-BL with $\Omega = \pi/T$. Denote
its energy by
\begin{equation}
 E_y = \int_{-\infty}^{\infty} |y(t)|^2 \, dt
 = \frac{1}{2\pi} \int_{-\pi/T}^{\pi/T} |Y(\omega)|^2 \, d\omega.
 \label{eq:Ey_def}
\end{equation}
Since $Y(\omega) = X(\omega)$ for $|\omega| \leq \pi/T$, the energy $E_y$
captures the portion of $x(t)$'s energy within the SI reconstruction band.
By the Cauchy-Schwarz inequality applied to the inverse Fourier integral, the energy bounds the amplitude~\cite{naaman_quant},
\begin{equation}
 \|y\|_\infty\leq \sqrt{\frac{E_y \Omega}{\pi}} = \sqrt{\frac{E_y}{T}},
 \label{eq:c_Ey}
\end{equation}
and we choose the conservative value $c=\sqrt{E_y/T}$ for the encoder design and the bounds below.

A uniform quantizer with $K = 2^B$ levels is applied to each inter-spike
interval $T_n$. From \eqref{eq:interval_bounds}, the intervals lie in
$[\Delta t_{\min}, \Delta t_{\max}]$ with
$\Delta t_{\min} = \kappa\delta/(b+c)$ and
$\Delta t_{\max} = \kappa\delta/(b-c)$.
Setting $b = \alpha c$ for fixed $\alpha > 1$, the quantization step is
\begin{equation}
 \Delta_{\mathrm{IF\text{-}TEM}}
 = \frac{\kappa\delta}{(\alpha+1)(\alpha-1)} \cdot \frac{2}{cK}
 = \frac{\kappa\delta}{(\alpha+1)(\alpha-1)} \cdot \frac{2}{K}
 \sqrt{\frac{T}{E_y}},
 \label{eq:quant_step_Ey}
\end{equation}
showing that $\Delta_{\mathrm{IF\text{-}TEM}}$ decreases as $E_y$ grows.

The energy $E_y$ is related to $x(t)$ through the baseband bounds
$\alpha_{\mathrm{b}},\beta_{\mathrm{b}}$ of Section~\ref{subsec:lpf}. Since
$Y(\omega)=H(\omega)A(e^{j\omega T})$ on the band and
$\sqrt{T\alpha_{\mathrm{b}}}\leq|H(\omega)|\leq\sqrt{T\beta_{\mathrm{b}}}$ a.e.\
there, Parseval's theorem gives
\begin{equation}
 \alpha_{\mathrm{b}}\,\|a\|_{\ell^2}^2 \leq E_y \leq
 \beta_{\mathrm{b}}\,\|a\|_{\ell^2}^2.
 \label{eq:Ey_bounds}
\end{equation}
For a baseband-confined generator these are the Riesz constants
$\alpha_0,\beta_0$; for a wideband generator they are the constants of the
surviving baseband copy, the difference being exactly the energy lost at the
prefilter. Theorem~\ref{thm:quant_si} uses the upper bound in \eqref{eq:Ey_bounds} to express the error bound through the coefficients.

\subsection{MSE Bound for BL Recovery of $y(t)$}
\label{subsec:quant_bl}

Denote the quantized inter-spike intervals by $\hat{T}_n = T_n + d_n$, where
$d_n$ is the quantization error. We model $\{d_n\}$ as i.i.d.\ uniform random
variables on
$[-\Delta_{\mathrm{IF\text{-}TEM}}/2,\,\Delta_{\mathrm{IF\text{-}TEM}}/2]$. The
signal reconstructed from the quantized intervals is denoted $\hat{y}(t)$, and
the error signal is $e(t) = y(t) - \hat{y}(t)$. This is an additive stochastic quantization-error model, not a claim that the errors of an undithered deterministic uniform quantizer are independent or uniform. Moreover, errors in intervals accumulate when firing times are reconstructed: with an exact reference time, $\hat t_n-t_n=\sum_{j=0}^{n-1}d_j$ for $n>0$. The imported BL bound is used only for the interval-error model and reconstruction covered by~\cite[Theorem~2]{naaman_quant}, not as a guarantee for arbitrary timestamp-jitter models. Following~\cite{naaman_quant},
define the normalized squared error
\begin{equation}
 \varepsilon^2 = \lim_{n\to\infty}
 \frac{1}{2n\,\Delta t_{\min}}
 \left\| e \cdot \mathbf{1}_{[-n\Delta t_{\min},\,n\Delta t_{\min}]} \right\|^2,
 \label{eq:mse_def}
\end{equation}
where the limit is taken pathwise. We assume that the limit in \eqref{eq:mse_def} exists almost surely and that $\mathbb{E}[\varepsilon^2]<\infty$, and set $\mathrm{MSE}_{\mathrm{BL}}=\mathbb{E}[\varepsilon^2]$. The SI propagation argument below is conditional on this BL-stage hypothesis; it does not establish existence of the limit.

\begin{lemma}
\label{lem:bl_mse}
Under the conditions of Theorem~\ref{thm:main}, with $b = \alpha c$,
$\alpha > 1$, and i.i.d.\ uniform quantization errors of step size
$\Delta_{\mathrm{IF\text{-}TEM}}$ as in \eqref{eq:quant_step_Ey}, the MSE
of the BL recovery of $y(t)$ satisfies
\begin{equation}
 \mathrm{MSE}_{\mathrm{BL}}
 < \frac{R}{(1-R)^2}
 \cdot \frac{\alpha+1}{\alpha-1}
 \cdot \frac{E_y}{3T}
 \cdot \frac{1}{2^{2B}},
 \label{eq:bl_mse_bound}
\end{equation}
where $R = \dfrac{\kappa\delta}{\alpha-1}\sqrt{\dfrac{1}{E_y T}} < 1$.
\end{lemma}

\begin{proof}
See Appendix~\ref{app:proofs} of the supplementary material.
\end{proof}

The factor $R$ in Lemma~\ref{lem:bl_mse} is not a new quantity: under the
parametrization $b = \alpha c$ it equals the convergence rate $r$ of the
iterative BL recovery \eqref{eq:convergence}, since $r =
\kappa\delta\Omega/((b-c)\pi) = (\kappa\delta/(\alpha-1))\sqrt{1/(E_y T)}$. The
bound \eqref{eq:bl_mse_bound} thus inherits the same $1/2^{2B}$ scaling known
from amplitude-quantized uniform sampling~\cite{naaman_quant}, with a prefactor
governed by the IF-TEM operating point; this comparison is made precise in
Section~\ref{subsec:quant_compare}.

\subsection{End-to-End MSE for SI Coefficient Recovery}
\label{subsec:quant_si}

The BL recovery error in $y(t)$ propagates through the SI correction filter
$1/H(\omega)$ to produce an error in the estimated coefficients $\hat{a}[n]$.
Proposition~\ref{prop:stability} bounds this propagation in terms of the
 baseband constants $\alpha_{\mathrm{b}}, \beta_{\mathrm{b}}$. Equation~\eqref{eq:mse_def} measures time-averaged squared error. A bound on this quantity does not imply a bound on the total error energy over an infinite record. We therefore bound the error per coefficient. For the window $W_n=[-n\Delta t_{\min},n\Delta t_{\min}]$, let $J_n=\{k:kT\in W_n\}$ and $M_n=|J_n|$, and let $\hat{a}_n$ be obtained from the record $\mathbf{1}_{W_n}\hat{y}$ by keeping its baseband spectrum and applying \eqref{eq:A_recovery} with $S\equiv1$. The following theorem combines
Lemma~\ref{lem:bl_mse} with Proposition~\ref{prop:stability} to give a
complete end-to-end bound.

\begin{theorem}
\label{thm:quant_si}
Let $\mathcal{V}$ be an SI space as in \eqref{eq:si} with generator $h(t)$
satisfying the baseband bounds
$0 < \alpha_{\mathrm{b}} \leq \beta_{\mathrm{b}} < \infty$ of
Section~\ref{subsec:lpf}, which for a baseband-confined generator are the Riesz
bounds \eqref{eq:riesz_single}. Let $x(t)\in\mathcal{V}$ with $0<E_y<\infty$, and assume the BL-stage error satisfies the almost-sure limit and integrability hypotheses stated after \eqref{eq:mse_def}.
Apply an LPF with cutoff $\pi/T$ to obtain $y(t)$, with energy $E_y$ and
amplitude bound $c = \sqrt{E_y/T}$ of \eqref{eq:c_Ey}. Encode $y(t)$ with an IF-TEM with
parameters $(b,\kappa,\delta)$, $b = \alpha c$ for fixed $\alpha > 1$,
satisfying \eqref{eq:density_condition}. Apply a $K$-level uniform quantizer
with $K = 2^B$ to the inter-spike intervals, with i.i.d.\ uniform
quantization errors.

Then the per-coefficient MSE

\[
 P_a=\mathbb{E}\Bigl[\limsup_{n\to\infty}\frac{1}{M_n}
 \sum_{k\in J_n}|\hat{a}_n[k]-a[k]|^2\Bigr]
\]
satisfies
\begin{equation}
 P_a\leq\frac{T}{\alpha_{\mathrm{b}}}\,\mathrm{MSE}_{\mathrm{BL}}
 <
 \frac{1}{ \alpha_{\mathrm{b}}}
 \cdot
 \frac{R}{(1-R)^2}
 \cdot
 \frac{\alpha+1}{\alpha-1}
 \cdot
 \frac{E_y}{3}
 \cdot
 \frac{1}{2^{2B}},
 \label{eq:si_coeff_mse}
\end{equation}
where $R = \dfrac{\kappa\delta}{\alpha-1}\sqrt{\dfrac{1}{E_y T}} < 1$.
If, moreover, $R\leq\frac13$, then, using $E_y \leq \beta_{\mathrm{b}} \|a\|_{\ell^2}^2$ from
\eqref{eq:Ey_bounds}, the bound can be expressed directly in terms of $x(t)$:
\begin{equation}
 P_a
 <
 \frac{ \beta_{\mathrm{b}}}{ \alpha_{\mathrm{b}}}
 \cdot
 \frac{R'}{(1-R')^2}
 \cdot
 \frac{\alpha+1}{\alpha-1}
 \cdot
 \frac{\|a\|_{\ell^2}^2}{3}
 \cdot
 \frac{1}{2^{2B}},
 \label{eq:si_coeff_mse_x}
\end{equation}
where $R' = \dfrac{\kappa\delta}{\alpha-1}\sqrt{\dfrac{1}{ \beta_{\mathrm{b}}\|a\|_{\ell^2}^2 T}}$.
\end{theorem}

\begin{proof}
See Appendix~\ref{app:proofs} of the supplementary material.
\end{proof}

\begin{remark}[Finite records and meaning of the error criterion]
For each finite window $W_n$, the same argument gives the nonasymptotic bound
\begin{equation}
 \frac{1}{M_n}\mathbb{E}\sum_{k\in J_n}|\hat a_n[k]-a[k]|^2
 \leq \frac{\mathbb{E}\|\mathbf{1}_{W_n}e\|_{L^2}^2
 +\|\mathbf{1}_{W_n^c}y\|_{L^2}^2}
 {\alpha_{\mathrm b}M_n}.
 \label{eq:finite_record_mse}
\end{equation}
Here $\hat a_n$ is a window-dependent estimator formed by zero extension, baseband projection, and SI inversion; it is not the restriction of a single infinite-record reconstruction. The asymptotic criterion $P_a$ is an absolute error per coefficient, not the finite-record coefficient NMSE in \eqref{eq:nmse_coef}. In particular, for $a\in\ell^2$, the average signal energy per coefficient vanishes as the record grows. No finite-record relative-error bound follows by dividing the asymptotic inequality by this vanishing average. A finite-record quantization theorem would additionally require a finite-window BL error bound, including boundary effects.
\end{remark}

The bound \eqref{eq:si_coeff_mse} decomposes into three interpretable factors.
\emph{Generator conditioning:} $1/\alpha_{\mathrm{b}}$ is set by the baseband
lower bound of $h(t)$, and a large $\alpha_{\mathrm{b}}$ limits amplification of
the BL recovery error by the correction filter. \emph{IF-TEM operating point:}
$R(1-R)^{-2}(\alpha+1)/(\alpha-1)$ depends on $(b,\kappa,\delta)$ through the
bias ratio $\alpha$ and the convergence rate $R$. \emph{Quantization
resolution:} the $1/2^{2B}$ factor is the standard decay with bit depth,
identical to that of a classical ADC~\cite{naaman_quant}. The dependence on
signal energy is governed by $g(E_y)=R(1-R)^{-2}E_y$, increasing on
$R\leq\frac13$ as shown in the proof: a higher-energy SI signal yields a smaller
quantization step through \eqref{eq:quant_step_Ey}, while the interval
measurement is scaled by $b=\alpha\sqrt{E_y/T}$ and the guaranteed firing rate
grows accordingly. The ratio of the bound to $E_y$ is invariant to a common scaling of the signal once
$\kappa\delta$ is re-set to hold the oversampling factor fixed, which is the
operating point used in Section~\ref{sec:sim}.

\subsection{Comparison with Amplitude Quantization in Uniform
Sampling}
\label{subsec:quant_compare}

To place Theorem~\ref{thm:quant_si} in context, consider the classical
benchmark: uniform samples of the same BL signal $y(t)$, taken at the rate
$T^{-1}$ and quantized in amplitude with $B$ bits over the dynamic range
$[-c,c]$. The amplitude quantization step is
$\Delta_{\mathrm{amp}}=2c/2^{B}$, and under the standard uniform-error model
the time-averaged quantization noise power is
\begin{equation}
 \mathrm{MSE}_{\mathrm{amp}}
 = \frac{\Delta_{\mathrm{amp}}^2}{12}
 = \frac{c^2}{3}\cdot\frac{1}{2^{2B}}
 = \frac{E_y}{3T}\cdot\frac{1}{2^{2B}},
 \label{eq:amp_benchmark}
\end{equation}
using $c=\sqrt{E_y/T}$ from \eqref{eq:c_Ey}. On the samples, the correction filter $T/H(\omega)$ has gain at most $\sqrt{T/\alpha_{\mathrm{b}}}$, so the per-coefficient error of this benchmark is at most $(T/\alpha_{\mathrm{b}})\,\mathrm{MSE}_{\mathrm{amp}}$, with the same factor $T/\alpha_{\mathrm{b}}$ as in \eqref{eq:si_coeff_mse}. The end-to-end comparison therefore reduces to comparing \eqref{eq:amp_benchmark} with the BL-stage bound \eqref{eq:bl_mse_bound}. The two differ exactly by the operating-point factor
\begin{equation}
 f(R,\alpha)=\frac{R}{(1-R)^2}\cdot\frac{\alpha+1}{\alpha-1},
 \label{eq:quant_ratio}
\end{equation}
that is, the time-encoding guarantee equals the classical amplitude-sampling
benchmark multiplied by $f(R,\alpha)$. Since both quantities are upper bounds,
the comparison is between guarantees. It shows that time encoding matches the
classical $6$~dB-per-bit decay, and that its guarantee is at least as strong as
the classical one whenever $f(R,\alpha)\leq 1$ at equal bit depth per measurement. Since $R$ decreases as the
firing rate grows, with $R=\alpha/(\mathrm{OF}(\alpha-1))$ under the
parametrization of Section~\ref{subsec:sim_setup}, where $\mathrm{OF}$ denotes
the oversampling factor, the factor can always be driven below one by moderate
oversampling: for $\alpha=3$, $f(R,3)=2R/(1-R)^2\leq 1$ for $R\leq
2-\sqrt{3}\approx 0.27$, that is, for $\mathrm{OF}\gtrsim 5.6$. This crossover,
together with the two reference settings, is illustrated in
Fig.~\ref{fig:ref_compare} of Section~\ref{subsec:sim_ref}. For fixed $\alpha$, the admissible parametrization requires $\mathrm{OF}>\alpha/(\alpha-1)$. Thus $R\uparrow1$ and $f\to\infty$ as $\mathrm{OF}\downarrow\alpha/(\alpha-1)$, not as $\mathrm{OF}\downarrow1$. Approaching the coefficient rate while retaining this sufficient guarantee requires increasing $\alpha$ as well.

The comparison is per measurement: at oversampling factor $\mathrm{OF}$ the
IF-TEM produces $\mathrm{OF}$ times more events per unit time than uniform
sampling at rate $T^{-1}$, and at the crossover $\mathrm{OF}\approx5.6$ it spends about $5.6$ times more bits per unit time at the same bit depth. At equal total bit rate the amplitude sampler may use $\mathrm{OF}\cdot B$ bits per sample, and the ratio of the two bounds becomes $f(R,\alpha)\,2^{2(\mathrm{OF}-1)B}$, which exceeds one for every $B\geq1$ at all operating points of Fig.~\ref{fig:ref_compare}(b). The advantage for $f(R,\alpha)\leq1$ therefore holds at equal bits per measurement, and, within this idealized quantization model, the uniform sampler has the smaller upper bound at the operating points considered. This compares bounds, not measured performance or an information-theoretic lower bound. The
quantities being quantized also differ in nature, the IF-TEM quantizing time
differences, which is precisely the low-power operation that motivates
time-based ADCs~\cite{mulleti2024power}.

The generator enters the end-to-end bound \eqref{eq:si_coeff_mse} through two
quantities only: the surviving energy
$E_y\in[\alpha_{\mathrm{b}}\|a\|_{\ell^2}^2,\beta_{\mathrm{b}}\|a\|_{\ell^2}^2]$
of \eqref{eq:Ey_bounds}, which sets the quantization step, and the conditioning
$1/\alpha_{\mathrm{b}}$, which sets the amplification. Well-conditioned,
baseband-concentrated generators are favorable on both counts, while generators
with in-band near-zeros are penalized through $1/\alpha_{\mathrm{b}}$. With the
mixer of Theorem~\ref{thm:robust} the same analysis applies with $H(\omega)$
replaced by $R(\omega)$ and $\alpha_{\mathrm{b}}$ by the corresponding lower
bound on $|R(\omega)|^2/T$, and with $E_y$, $c$, the quantization step and $R$ evaluated for the mixed waveform, since the mixer changes all of them. The $1/2^{2B}$ decay is unaffected.

\section{Evaluation Results}
\label{sec:sim}

The experiments parallel the modulo-sampling study of~\cite{kvich2024modulo}
with the modulo sampler replaced by the IF-TEM: the analog observable is the set
of firing times, equivalently the inter-spike intervals, and the inner BL stage
uses the iterative recovery \eqref{eq:iteration} rather than a modulo unfolding.
We establish the setup and error metrics (Section~\ref{subsec:sim_setup}),
demonstrate noise-free recovery across four signal classes
(Appendix~\ref{subsec:sim_recovery} of the supplementary material), study noise resilience for a
generator without in-band zeros (Appendix~\ref{subsec:sim_noise}) and the role of the mixer for
a Riesz generator with in-band zeros
(Section~\ref{subsec:sim_mixer}), compare against direct SI inversion and the
crossing-based encoder of~\cite{gontier2014sampling}
(Section~\ref{subsec:sim_compare}), and test the bit-depth scaling of
Theorem~\ref{thm:quant_si} (Section~\ref{subsec:sim_quant}).

\subsection{Experimental Setup and Performance Metrics}
\label{subsec:sim_setup}

Throughout, continuous-time signals are represented on a dense uniform grid and
the SI period is fixed to $T = 1$. An SI signal is synthesized as $x(t) =
\sum_n a[n]\,h(t-nT)$ with i.i.d.\ coefficients $a[n]$, low-pass filtered to
$y(t) = \mathrm{LPF}_{\pi/T}\{x(t)\}$, or to
$y(t)=\mathrm{LPF}_{\pi/T}\{p(t)x(t)\}$ when a mixer is used, and encoded by
the IF-TEM. The operating point is set from the signal amplitude bound: we take
$c = 1.02\,\max_t|y(t)|$, bias $b = \alpha c$ with $\alpha = 3$, scaling
$\kappa = 1$, and threshold $\delta$ chosen so that the mean firing rate
$b/(\kappa\delta)$ equals an oversampling factor (OF) times the coefficient rate
$T^{-1}$. With this parametrization the convergence and density factor of
\eqref{eq:convergence} becomes
\begin{equation}
 r \;=\; \frac{\kappa\delta\,\Omega}{(b-c)\pi}
 \;=\; \frac{\alpha}{\mathrm{OF}\,(\alpha-1)} \;<\; 1,
 \label{eq:sim_r}
\end{equation}
so that condition \eqref{eq:density_condition} holds for all reported OF
values, which is equivalent to $\mathrm{OF}>\alpha/(\alpha-1)$. The BL recovery
\eqref{eq:iteration} is run to convergence using the sine-integral
representation of the operator \eqref{eq:rec_operator}, and the SI coefficients
are then extracted by the correction filter $1/H(\omega)$ for a single
generator, $1/R(\omega)$ when a mixer is used, and
$\mathbf{M}^\dagger(\omega)$ for multiple generators.

To emulate front-end noise, additive white Gaussian noise is added to the
integral measurements $y_n$ of \eqref{eq:measurement_y_n}, the quantity consumed
by the recovery operator and by the error term $\varepsilon$ of
Proposition~\ref{prop:stability}; the signal-to-noise ratio (SNR) is defined on
$\{y_n\}$. We report two NMSE metrics, matching~\cite{kvich2024modulo}. The
\emph{BL error} measures the discrepancy of the recovered BL samples,
\begin{equation}
 \mathrm{NMSE}_{\mathrm{BL}}
 = \frac{\sum_n |y[nT_s] - \hat{y}[nT_s]|^2}{\sum_n |y[nT_s]|^2},
 \label{eq:nmse_bl}
\end{equation}
and the \emph{coefficient error} measures the precision of the recovered SI
coefficients after the correction filter,
\begin{equation}
 \mathrm{NMSE}_{\mathrm{coef}}
 = \frac{\|a - \hat{a}\|_2^2}{\|a\|_2^2}.
 \label{eq:nmse_coef}
\end{equation}
Both are reported in decibels. Separating them isolates the inner BL stage from
the SI inversion, exposing the error amplification analyzed in
Propositions~\ref{prop:stability} and~\ref{prop:stability_multi}. Unless stated
otherwise, each SNR point is averaged over $60$ Monte Carlo trials.

\subsection{Riesz Generators with In-Band Zeros}
\label{subsec:sim_mixer}

To probe the regime in which the mixer is essential, we use a scaled
first-order B-spline, $h(t) = \beta^{1}(t/2.5)$, a Riesz
generator whose frequency response $H(\omega)$ has zeros inside
$[-\pi/T,\pi/T]$. By Theorem~\ref{thm:main} the signal is still identifiable
wherever $H(\omega)\neq 0$, but the correction filter $1/H(\omega)$ is ill
conditioned: the BL recovery error of Proposition~\ref{prop:stability} is
amplified at the near-zero frequencies, severely degrading the coefficient
error.

\begin{figure}[t!]
 \centering
 \includegraphics[width=0.6\linewidth]{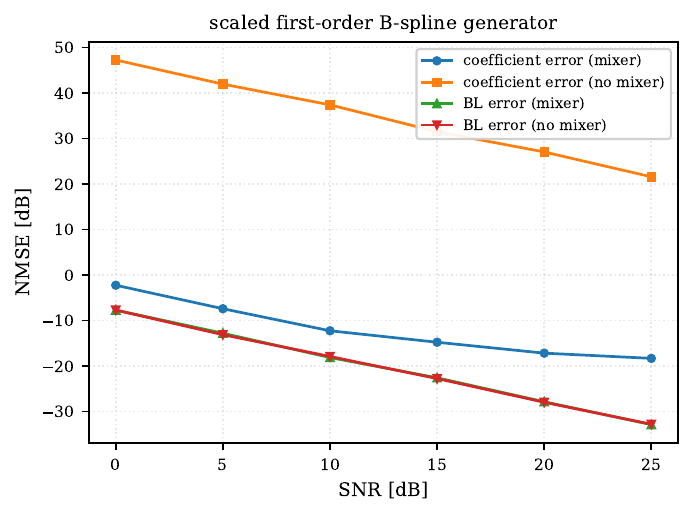}
 \caption{NMSE in dB versus SNR for a scaled first-order B-spline generator,
 whose $H(\omega)$ has zeros inside the reconstruction band. Error metrics
 follow \eqref{eq:nmse_bl}-\eqref{eq:nmse_coef}. Without a mixer the
 correction filter is unstable and the coefficient error diverges from the
 BL error; mixing restores a well-conditioned response and
 reduces the coefficient NMSE by approximately $40$--$50$~dB over the plotted
 SNR range.}
 \label{fig:snr_spline}
\end{figure}

The mixer
$p(t)=1+1000\cos(2\pi t/T)+1000\cos(4\pi t/T)$
is selected so that $R(\omega)$ in
\eqref{eq:R_omega} has no in-band zeros. It lowers the
coefficient NMSE by approximately $40$--$50$~dB over the
plotted SNR range (Fig.~\ref{fig:snr_spline}). Without mixing,
the coefficient error is much larger than the BL error,
consistent with the unbounded inverse response at the
generator's in-band zeros. With mixing, the coefficient NMSE
reaches approximately $-18$~dB at the highest reported SNR
and remains above the BL error. This is the IF-TEM counterpart
of \cite[Fig.~8]{kvich2024modulo} and illustrates the stability improvement
obtained by replacing the singular response $H(\omega)$ with a mixed
response $R(\omega)$ bounded away from zero.

The mechanism behind this behavior, the shape of the effective filter $R(\omega)$ and of the correction filter $1/R(\omega)$, is illustrated in Appendix~\ref{app:corr_filter} of the supplementary material.

\emph{Matched prefilter.} The same spline also illustrates Corollary~\ref{cor:riesz}, with $g(t)=h(-t)$ and unit mixer weights on $\mathcal{K}=\{-m,\ldots,m\}$. For $m=0$ the lower bound $\alpha_{\mathcal{K}}$ vanishes, since the baseband copy has in-band zeros. The adjacent copies have no common zero, so one harmonic on each side already gives $\alpha_{\mathcal{K}}=0.009211$, within $0.5\%$ of the lower Riesz bound $\alpha_0\approx0.009257$, as Corollary~\ref{cor:riesz} predicts. We periodized the spline over $31T$, encoded a zero-mean Gaussian coefficient record at $T=1$ and $\mathrm{OF}=5$ with $b=3c$, where $c$ is the sum of the magnitudes of the Fourier-series coefficients of $y(t)$, and recovered the BL Fourier modes from the $155$ interval integrals by least squares. For $m=1$ the coefficient NMSE is below $-200$~dB, and the correction filter amplifies a perturbation at the least favorable frequency of this grid by $108.5$, within the bound $1/(\sqrt{T}\alpha_{\mathcal{K}})\approx108.6$ of \eqref{eq:riesz_stability}.\par

\subsection{Comparison with Direct Time Encoding of SI Signals}
\label{subsec:sim_compare}

\begin{table*}[t!]
 \centering
 \caption{Median coefficient NMSE (dB) over 40 trials versus timestamp jitter. Left: direct SI inversion of IF-TEM measurements~\cite{gontier2014sampling}, which receives the original SI signal, and the proposed front end, with 73 intervals per record in both routes. Right: crossing-TEM~\cite{gontier2014sampling} and the proposed IF-TEM with a common low-pass input and approximately 280 events per record. Bold marks the lower error within each pair.}
 \label{tab:compare_gv}
 \small\setlength{\tabcolsep}{5.5pt}
 \begin{tabular}{c rr rr @{\hspace{12pt}} rr rr}
 \toprule
 & \multicolumn{4}{c}{Direct SI inversion: 73 intervals} & \multicolumn{4}{c}{Common low-pass input: $\sim$280 events} \\
 \cmidrule(lr){2-5}\cmidrule(lr){6-9}
 & \multicolumn{2}{c}{Cubic B-spline} & \multicolumn{2}{c}{Band-pass} & \multicolumn{2}{c}{Sinc} & \multicolumn{2}{c}{Cubic B-spline} \\
 $\sigma_t/T$ & Proposed & Direct & Proposed & Direct & Proposed & Crossing & Proposed & Crossing \\
 \midrule
 $10^{-4}$ & $-51.1$ & $\mathbf{-51.8}$ & $\mathbf{-60.5}$ & $-35.0$ & $-43.2$ & $\mathbf{-49.5}$ & $-36.3$ & $\mathbf{-41.5}$ \\
 $3\times10^{-4}$ & $-41.6$ & $\mathbf{-42.3}$ & $\mathbf{-50.9}$ & $-25.4$ & $-43.5$ & $\mathbf{-44.2}$ & $-35.8$ & $\mathbf{-38.2}$ \\
 $10^{-3}$ & $-31.1$ & $\mathbf{-31.8}$ & $\mathbf{-40.5}$ & $-15.0$ & $\mathbf{-40.8}$ & $-34.5$ & $\mathbf{-34.4}$ & $-30.4$ \\
 $3\times10^{-3}$ & $-21.5$ & $\mathbf{-22.3}$ & $\mathbf{-30.9}$ & $-5.4$ & $\mathbf{-34.2}$ & $-25.5$ & $\mathbf{-26.4}$ & $-21.6$ \\
 $10^{-2}$ & $-10.8$ & $\mathbf{-11.8}$ & $\mathbf{-20.2}$ & $5.3$ & $\mathbf{-24.7}$ & $-14.9$ & $\mathbf{-17.6}$ & $-11.9$ \\
 \bottomrule
 \end{tabular}
\end{table*}

\emph{Direct IF-TEM inversion.} The most direct comparison keeps the IF-TEM and changes only the route from firing times to coefficients. The direct encoder receives $x(t)$ without an LPF, and its decoder solves the IF-TEM measurement equations of~\cite{gontier2014sampling} in the SI space by least squares, $\mathbf{D}\mathbf{a}=\kappa\delta\mathbf{1}-b\,\Delta\mathbf{t}$ with $[\mathbf{D}]_{i,k}=\int_{t_i}^{t_{i+1}}h(u-kT)\,du$. Our encoder receives the output of the analog front end, and its decoder recovers the BL Fourier modes from the interval integrals before the SI correction. We used $T=1$, a periodic record of $31$ zero-mean Gaussian coefficients and $73$ intervals per record, about $2.4$ per period, for both encoders. Both share $b=3c$, where $c$ bounds both encoder inputs, $\kappa=1$ and $\delta=31b/73$, and the same Gaussian timing perturbations in each of $40$ paired trials; both decoders use the perturbed firing times. For a cubic B-spline our front end is the LPF alone (Theorem~\ref{thm:main}). For the band-pass generator $h(t)=2\cos(6\pi t)\,\mathrm{sinc}(\pi t)$ it is the architecture of Corollary~\ref{cor:riesz} with $\mathcal{K}=\{-3,3\}$: $H(\omega)$ equals one on its support, so $g(t)=h(-t)$ passes $x(t)$ unchanged, $p(t)=2\cos(6\pi t)$, and $R_{\mathcal{K}}(\omega)=2$ on the band, so that $\alpha_{\mathcal{K}}=\alpha_0$, whereas the LPF alone retains nothing of $x(t)$. With exact timestamps, median NMSE was below $-200$~dB for both routes and generators. Under jitter, direct inversion is $0.7$--$1.0$~dB more accurate for the cubic spline, and the proposed front end is $25.5$~dB more accurate for the band-pass generator at $\sigma_t=10^{-3}T$, as summarized in Table~\ref{tab:compare_gv}.\par

\emph{Crossing-based encoding.} We compare the proposed pipeline against the crossing-based time encoding and
reconstruction of~\cite{gontier2014sampling}, complementing the conceptual
comparison of Section~\ref{subsec:gv_relation}. Both systems are placed on an
equal footing: each is preceded by the anti-alias LPF, both operate at the same
average number of events, and the encoders differ only in their sampling
mechanism. This compares encoders on a common BL input rather than full approaches, since the method of~\cite{gontier2014sampling} does not need the LPF. The proposed encoder is the biased IF-TEM, whose information is
carried by the inter-spike intervals \eqref{eq:measurement_y_n}. The reference
encoder is the crossing-TEM of~\cite{gontier2014sampling}: events are the
instants at which the input meets the test function $\Phi(t)=A\cos(2\pi t/P)$,
the decoder reads the amplitudes $\Phi(t_i)$. Because this experiment includes an LPF, its generator-sample least-squares estimator uses the effective generator $h_{\mathrm{LP}}=\mathrm{LPF}_{\pi/T}\{h\}$: $\hat{\mathbf a}=\mathbf D_{\mathrm{cross}}^{\dagger}\mathbf f$, with $[\mathbf D_{\mathrm{cross}}]_{i,k}=h_{\mathrm{LP}}(t_i-kT)$ and $f_i=\Phi(t_i)$.

\emph{Simulation conditions.} For both generators the input is a random SI
signal of $K=32$ coefficients drawn i.i.d.\ $\mathcal{N}(0,1)$ and BL by the
anti-alias LPF. The proposed IF-TEM uses the bias-to-amplitude margin
$\alpha=b/c=2.5$ and integrator constant $\kappa=1$; the threshold $\delta$ is
fixed so that the guaranteed firing rate $(b-c)/(\kappa\delta)$ equals
$\mathrm{OF}/T$ with oversampling factor $\mathrm{OF}=5$, that is,
$\delta=(\alpha-1)\,c\,T/(\kappa\,\mathrm{OF})$, which satisfies the
identifiability condition $\kappa\delta/(b-c)<T$ of Theorem~\ref{thm:main} with
margin. This yields $N\approx 280$ firing instants $\{t_n\}$ per realization.
For the crossing-TEM the amplitude is set to the smallest admissible value
$A=1.05\,\|y\|_\infty$, which is the most favorable conditioning for that
decoder, and the period $P$ is set so that its crossing count matches the
IF-TEM event count ($N\approx 280$). Coefficient NMSE is reported as the median
over $40$ Monte Carlo trials.

\emph{Input-signal noise.} With additive in-band noise the two methods are
comparable across the SNR range: the coefficient NMSE agrees to within about
$1$~dB up to moderate SNR, and the crossing-TEM is marginally better at high
SNR, where its least-squares inversion over the full event set averages the
measurement noise effectively. This regime does not separate the two
approaches.

\emph{Timing error on the recorded events.} The two methods are distinguished
by their sensitivity to perturbations of the event times themselves, which is
the quantization and jitter impairment that dominates clockless, low-power
time-based hardware, and the regime analyzed
in~\cite[Sec.~6]{gontier2014sampling}. The event times are perturbed by
zero-mean Gaussian noise of standard deviation $\sigma_t$, in units of $T$; all
other settings are as above. When the event times are nearly exact
($\sigma_t \lesssim 5\times10^{-4}T$), the crossing-TEM is the more accurate of
the two by a few dB, consistent with the stability its reconstruction is proven
to possess. At $\sigma_t/T$ from $10^{-3}$ to $10^{-2}$, the proposed IF-TEM yields $6$--$10$~dB lower coefficient NMSE for the sinc and $4$--$6$~dB lower NMSE for the spline (Table~\ref{tab:compare_gv}, right).

The measurement perturbations alone do not determine the reconstruction error. If $\hat t_i=t_i+\eta_i$, the IF-TEM measurement computed from the perturbed times changes exactly by $-b(\eta_{i+1}-\eta_i)$, whereas a crossing amplitude changes by $\Phi'(t_i)\eta_i+O(\eta_i^2)$. In both methods the reconstruction matrix also changes because its entries depend on the recorded times. For direct IF-TEM inversion,
\[
 \Delta D_{i,k}=h(t_{i+1}-kT)\eta_{i+1}
              -h(t_i-kT)\eta_i+O(\|\boldsymbol\eta\|^2),
\]
when the generator is sufficiently smooth. Timing sensitivity therefore depends on the input and test-function slopes, the perturbed reconstruction operator, and its conditioning. The observed differences in Table~\ref{tab:compare_gv} are empirical results for the stated operating points, not a general consequence of replacing a slope factor by the bias $b$.

\subsection{Quantization of the Inter-Spike Intervals}
\label{subsec:sim_quant}

Finally, we test the bit-depth scaling of the quantization analysis of
Section~\ref{sec:quant}. The inter-spike intervals $T_n$ are uniformly
quantized with $K=2^B$ levels over $[\Delta t_{\min},\Delta t_{\max}]$, and the
coefficients are recovered from the quantized intervals. The coefficient NMSE
versus the bit depth $B$ is reported in Fig.~\ref{fig:quant} for the four
signal classes of Appendix~\ref{subsec:sim_recovery}. As in Section~\ref{subsec:sim_setup}, the encoder and the quantizer use $c=1.02\max_t|y(t)|$, whereas Theorem~\ref{thm:quant_si} uses the conservative $c=\sqrt{E_y/T}$ of \eqref{eq:c_Ey}; the experiment therefore tests the $1/2^{2B}$ scaling of the bound and not its constant.

For every generator the curves decay at approximately $6$~dB per bit, matching
the $1/2^{2B}$ factor predicted by Theorem~\ref{thm:quant_si} and consistent
with classical amplitude-quantized ADCs~\cite{naaman_quant}, in agreement with
Section~\ref{subsec:quant_compare}. At high bit depth the curves flatten to a floor set
by the finite-iteration BL recovery and the finite-length correction filter
rather than by quantization, so the $1/2^{2B}$ trend holds over the
low-to-moderate bit-depth range that motivates time-based ADCs.

\begin{figure}[t!]
 \centering
 \includegraphics[width=0.6\linewidth]{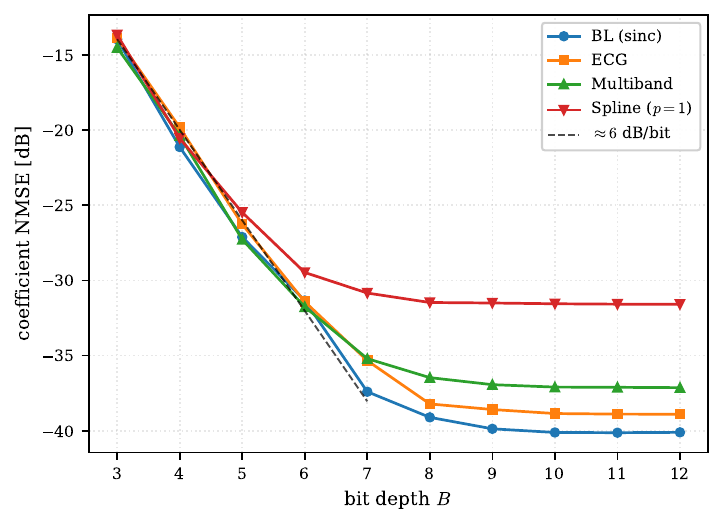}
 \caption{Coefficient NMSE versus bit depth $B$ for uniform quantization of
 the IF-TEM inter-spike intervals, for the four signal classes of
 Fig.~\ref{fig:recovery}. The common $\approx 6$~dB/bit slope matches the
 $1/2^{2B}$ scaling of Theorem~\ref{thm:quant_si}.}
 \label{fig:quant}
\end{figure}
\begin{figure*}[t!]
 \centering
 \includegraphics[width=.7\textwidth]{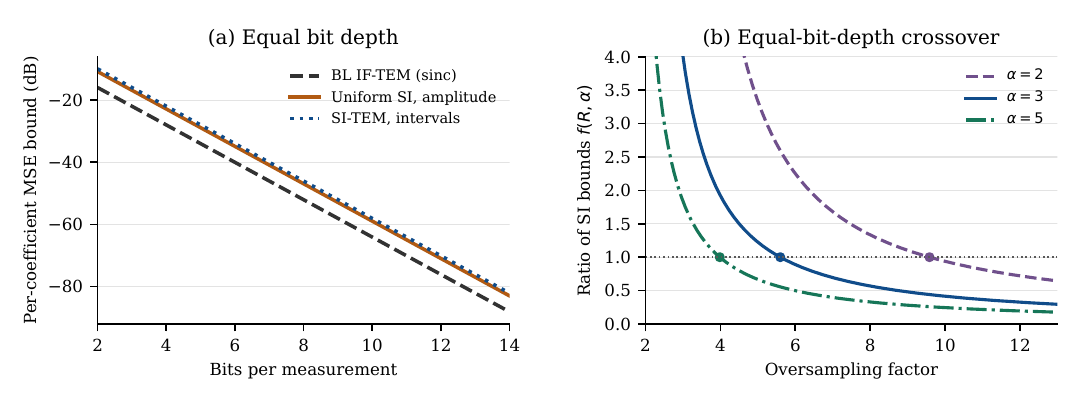}
 \caption{Per-coefficient MSE guarantees at equal bits per measurement, from the closed-form bounds of Section~\ref{sec:quant}. (a)~Bounds versus bit depth for IF-TEM sampling of BL signals (sinc generator), uniform SI sampling with amplitude quantization, and the proposed SI-TEM; $T=1$, $E_y/T=1$, $\alpha=3$, $\mathrm{OF}=5$, and $T/\alpha_{\mathrm{b}}=4$ for both SI curves. (b)~Ratio $f(R,\alpha)$ of the SI-TEM and uniform SI bounds versus the oversampling factor; markers indicate $f=1$.}
 \label{fig:ref_compare}
\end{figure*}
\subsection{Comparison with the Two Reference Settings}
\label{subsec:sim_ref}

We close by placing the proposed SI-TEM guarantee side by side with the two
reference settings, uniform sampling of SI signals (Section~\ref{subsec:si}) and
IF-TEM sampling of BL signals (Section~\ref{subsec:rec_IF}).
Fig.~\ref{fig:ref_compare} plots the closed-form per-coefficient MSE guarantees of
Section~\ref{sec:quant} rather than Monte Carlo points, so the comparison
reflects the theory directly.

All three guarantees share the $1/2^{2B}$ decay of Theorem~\ref{thm:quant_si},
the classical $6$~dB/bit slope of amplitude-quantized ADCs~\cite{naaman_quant},
and differ only by two interpretable offsets, as shown in panel~(a) at a common
operating point ($\alpha=3$, $\mathrm{OF}=5$, $E_y/T=1$, and $T/\alpha_{\mathrm{b}}=4$ for both SI curves). Read against the BL-TEM reference
the proposed curve isolates the factor $1/\alpha_{\mathrm{b}}$: for the sinc
generator $\alpha_{\mathrm{b}}=\alpha_0=T$ and the two coincide, recovering the
bound of~\cite{lazar2004perfect,naaman_quant} exactly, whereas a general
generator is penalized by $\sqrt{T/\alpha_{\mathrm{b}}}$ in amplitude, $6$~dB in MSE in panel~(a), the new
ingredient contributed by the SI inversion (Proposition~\ref{prop:stability}).
Read against the uniform-SI reference it isolates the operating-point factor
$f(R,\alpha)$ of \eqref{eq:quant_ratio}, the price of recovering the uniform
samples from the firing times rather than acquiring them with a clock.

That second comparison is made quantitative across operating points in
panel~(b), which plots $f(R,\alpha)$ versus the oversampling factor using
$R=\alpha/(\mathrm{OF}(\alpha-1))$. The factor falls below one, so that the
time-encoding guarantee becomes at least as strong as the amplitude-sampling
benchmark at equal bits per measurement, once the firing rate exceeds a modest multiple of $T^{-1}$, for
$\alpha=3$ at $\mathrm{OF}\gtrsim5.6$, and it grows without bound as $\mathrm{OF}$ decreases to $\alpha/(\alpha-1)$ for fixed $\alpha$, reflecting the slow convergence at the boundary of the sufficient density condition. At equal total bit rate the idealized uniform-sampling benchmark has the smaller upper bound at the operating points considered (Section~\ref{subsec:quant_compare}); this is not a universal performance ordering. The figure thus shows in one view what the SI-TEM inherits from each
reference setting and what is new, in line with
Section~\ref{subsec:quant_compare}.

\section{Conclusion}
\label{sec:conc}

We developed an IF-TEM framework that maps SI signals to BL waveforms, yielding a sufficient event-density condition set by the coefficient rate. Periodic mixing restores spectral copies lost by low-pass filtering. With a matched prefilter, a positive lower bound on the truncated Riesz sum ensures stable inversion. For multiple generators, full column rank almost everywhere ensures identifiability, while a uniform singular-value bound ensures stability. The stochastic interval-quantization analysis gives an asymptotic per-coefficient MSE bound with six-decibels-per-bit decay.

In the experiments, mixing restores stable recovery for a scaled B-spline with in-band zeros. At equal event counts and under timing error, the matched front end lowers coefficient error by about 25~dB relative to direct SI inversion of IF-TEM firing times for a band-pass generator; direct inversion is $0.7$--$1.0$~dB more accurate for the cubic spline at the reported jitter levels. These results show the conditioning benefit of analog spectral folding for generators whose baseband copy is weak or absent. The analysis assumes ideal analog components. A next step is to carry the analysis through measured nonideal filters and mixers and quantify the resulting accuracy and hardware-cost tradeoffs.

\bibliographystyle{IEEEtran}
\bibliography{refs}

@article{kvich2024modulo,
  title={Modulo Sampling in Shift-Invariant Spaces: Recovery and Stability Enhancement},
  author={Kvich, Yhonatan and Eldar, Yonina C},
  journal={arXiv preprint arXiv:2406.10929},
  year={2024}
}

@article{naaman2023hardware,
  title={{Hardware prototype of a time-encoding sub-Nyquist ADC}},
  author={Naaman, Hila and Glazer, Nimrod and Namer, Moshe and Bilik, Daniel and Savariego, Shlomi and Eldar, Yonina C},
  journal={IEEE Transactions on Instrumentation and Measurement},
  year={2024},
  publisher={IEEE}
}

@inproceedings{rastogi2011integrate,
  title={{Integrate and fire circuit as an ADC replacement}},
  author={{M. Rastogi, A. Singh Alvarado, J. G. Harris, and J. C. Príncipe}},
  booktitle={Proc. IEEE Int. Symp. Circuits Syst.},
  pages={2421--2424},
  year={2011}
}

@article{ryu2021time,
  title={{A Time-Based Pipelined {ADC} Using Integrate-and-Fire Multiplying-{DAC}}},
  author={{S. Ryu, C. Y. Park, W. Kim, S. Son, and J. Kim}},
  journal={IEEE Trans. Circuits Syst. I},
  volume={68},
  number={7},
  pages={2876-2889},
  year={2021}
}

@article{lazar2004perfect,
  title={Perfect recovery and sensitivity analysis of time encoded bandlimited signals},
  author={A. A. Lazar and L. T. T{\'o}th},
  journal={IEEE Trans. Circuits Syst. I},
  volume={51},
  number={10},
  pages={2060--2073},
  year={2004}
}

@article{lazar2004time,
  title={Time encoding with an integrate-and-fire neuron with a refractory period},
  author={A. A. Lazar},
  journal={Neurocomputing},
  volume={58},
  pages={53--58},
  year={2004}
}

@article{adam2020sampling,
  title={Sampling and reconstruction of bandlimited signals with multi-channel time encoding},
  author={{K. Adam, A. Scholefield, and M. Vetterli}},
  journal={IEEE Trans. Signal Process.},
   volume={68},
  number={},
  pages={1105-1119},
  year={2020}
}

@article{nyquist1928certain,
  title={Certain topics in telegraph transmission theory},
  author={Nyquist, H.},
  journal={Trans. Amer. Inst. Electr. Engineers},
  volume={47},
  number={2},
  pages={617--644},
  year={1928}
}

@article{aldroubi2001nonuniform,
  title={Nonuniform sampling and reconstruction in shift-invariant spaces},
  author={Aldroubi, Akram and Gr{\"o}chenig, Karlheinz},
  journal={SIAM review},
  volume={43},
  number={4},
  pages={585--620},
  year={2001},
  publisher={SIAM}
}

@article{gontier2014sampling,
  title={{Sampling based on timing: Time encoding machines on shift-invariant subspaces}},
  author={Gontier, D. and Vetterli, M.},
  journal={Applied and Comput. Harmonic Anal.},
  volume={36},
  number={1},
  pages={63--78},
  year={2014}
}

@article{alexandru2019reconstructing,
  title={Reconstructing classes of non-bandlimited signals from time encoded information},
  author={Alexandru, R. and Dragotti, P. L.},
  journal={IEEE Trans. Signal Process.},
  volume={68},
  pages={747--763},
  year={2019}
}

@inproceedings{alvarado2011integrate,
  title={{The integrate-and-fire sampler: A special type of asynchronous $\Sigma$-$\Delta$ modulator}},
  author={Alvarado, Alexander Singh and Rastogi, Manu and Harris, John G and Principe, Jose C},
  booktitle={2011 IEEE International Symposium of Circuits and Systems (ISCAS)},
  pages={2031--2034},
  year={2011},
  organization={IEEE}
}

@article{unser2000sampling,
  title={Sampling-50 years after Shannon},
  author={Unser, Michael},
  journal={Proc. IEEE},
  volume={88},
  number={4},
  pages={569--587},
  year={2000},
  publisher={IEEE}
}

@article{shannon1949communication,
  title={Communication in the presence of noise},
  author={Shannon, Claude Elwood},
  journal={Proceedings of the IRE},
  volume={37},
  number={1},
  pages={10--21},
  year={1949},
  publisher={IEEE}
}

@book{eldar2015sampling,
  title={{Sampling theory: Beyond bandlimited systems}},
  author={Eldar, Y. C},
  publisher={Cambridge University Press},
  year={2015}
}

@article{eldar2009compressed,
  title={Compressed sensing of analog signals in shift-invariant spaces},
  author={Eldar, Yonina C},
  journal={IEEE Transactions on Signal Processing},
  volume={57},
  number={8},
  pages={2986--2997},
  year={2009},
  publisher={IEEE}
}

@article{christensen2004oblique,
  title={Oblique dual frames and shift-invariant spaces},
  author={Christensen, Ole and Eldar, Yonina C},
  journal={Applied and Comput. Harmonic Anal.},
  volume={17},
  number={1},
  pages={48--68},
  year={2004},
  publisher={Elsevier}
}

@article{deboor1994structure,
  title={The structure of finitely generated shift-invariant spaces in L2 (Rd)},
  author={Deboor, Carl and DeVore, Ronald A and Ron, Amos},
  journal={Journal of Functional Analysis},
  volume={119},
  number={1},
  pages={37--78},
  year={1994},
  publisher={Elsevier}
}

@ARTICLE{neuromorphic_computer,  
title={{Low-Power Neuromorphic Hardware for Signal Processing Applications: A Review of Architectural and System-Level Design Approaches}},
author={{B. Rajendran, A. Sebastian, M. Schmuker, N. Srinivasa, and E. Eleftheriou}}, 
journal={IEEE Signal Process. Mag.},  volume={36},  number={6},  pages={97-110},year={2019}}

@ARTICLE{camera1,  
title={{Contour Motion Estimation for Asynchronous Event-Driven Cameras}}, 
author={{F. Barranco, C. Fermüller, and Y. Aloimonos}},  
journal={Proc. IEEE},  
volume={102},
number={10},
pages={1537-1556},
year={2014}}

@article{feichtinger2012approximate,
  title={Approximate reconstruction of bandlimited functions for the integrate and fire sampler},
  author={Feichtinger, Hans G and Pr{\'\i}ncipe, Jos{\'e} C and Romero, Jos{\'e} Luis and Alvarado, Alexander Singh and Velasco, Gino Angelo},
  journal={Adv. Comput. Math.},
  volume={36},
  number={1},
  pages={67--78},
  year={2012},
  publisher={Springer}
}

@book{miskowicz2018event,
  title={Event-based control and signal processing},
  author={M. Mi{\'s}kowicz},
  year={2018},
  publisher={CRC press}
}

@article{gray1998quantization,
  title={Quantization},
  author={Gray, Robert M. and Neuhoff, David L.},
  journal={IEEE Trans. on information theory},
  volume={44},
  number={6},
  pages={2325--2383},
  year={1998},
  publisher={IEEE}
}

@inproceedings{naaman2021time,
  title={{Time-Based Quantization for FRI and Bandlimited Signals}},
  author={Hila Naaman and Satish Mulleti and Y. C Eldar and Alejandro Cohen},
  booktitle={2022 30th European Signal Processing Conference (EUSIPCO)},
  pages={2241--2245},
 year={2022},
  organization={IEEE}
}

@article{naaman2021fri,
  title={{FRI-TEM: Time encoding sampling of finite-rate-of-innovation signals}},
  author={Naaman, Hila and Mulleti, Satish and Eldar, Yonina C},
  journal={IEEE Transactions on Signal Processing},
  volume={70},
  pages={2267--2279},
  year={2022},
  publisher={IEEE}
}

@ARTICLE{naaman_quant,
  author = {Hila Naaman and Neil Irwin Bernardo and Alejandro Cohen and Yonina C. Eldar},
  journal={IEEE Access}, 
  title={{Time Encoding Quantization of Bandlimited and Finite-Rate-of-Innovation Signals}}, 
  year={2025},
  volume={13},
  number={},
  pages={184745-184761},
  doi={10.1109/ACCESS.2025.3626162}}

@incollection{mulleti2024power,
  title={Power-Aware Analog to Digital Converters},
  author={Mulleti, Satish and Bhandari, Ayush and Eldar, Yonina C},
  booktitle={Sampling, Approximation, and Signal Analysis: Harmonic Analysis in the Spirit of J. Rowland Higgins},
  pages={415--452},
  year={2024},
  publisher={Springer}
}

@inproceedings{soundarapandian2024multigenerator,
  title={{On time-encoded sampling for multigenerator shift invariant spaces}},
  author={Soundarapandian, Roshaan and Budkuley, Amitalok J. and Rini, Stefano},
  booktitle={Proc. IEEE Int. Conf. Acoust., Speech, Signal Process. (ICASSP)},
  pages={9741--9745},
  year={2024}
}

@article{kamath2023neuromorphic,
  title={{Neuromorphic sampling of signals in shift-invariant spaces}},
  author={Kamath, Abijith Jagannath and Seelamantula, Chandra Sekhar},
  journal={arXiv preprint arXiv:2306.05103},
  year={2023}
}

@article{mishali2011subnyquist,
  title={{Sub-Nyquist sampling: Bridging theory and practice}},
  author={Mishali, M. and Eldar, Y. C.},
  journal={IEEE Signal Process. Mag.},
  volume={28},
  number={6},
  pages={98--124},
  year={2011}
}

@article{mishali2010theory,
  title={{From theory to practice: Sub-Nyquist sampling of sparse wideband analog signals}},
  author={Mishali, M. and Eldar, Y. C.},
  journal={IEEE J. Sel. Topics Signal Process.},
  volume={4},
  number={2},
  pages={375--391},
  year={2010},
  publisher={IEEE}
}

@article{gedalyahu2011multichannel,
  author  = {Gedalyahu, Kfir and Tur, Ronen and Eldar, Yonina C.},
  title   = {Multichannel Sampling of Pulse Streams at the Rate of Innovation},
  journal = {IEEE Trans. Signal Process.},
  volume  = {59}, number = {4}, pages = {1491--1504}, year = {2011}}

\clearpage
\setcounter{page}{1}
\twocolumn[{%
\begin{center}
{\Huge Supplementary Material for\\[1mm]
``Time-Based Sampling in Shift-Invariant Spaces''}\\[4mm]
{\large Hila~Naaman, Yhonatan~Kvich, and Yonina~C.~Eldar}
\end{center}
\vspace{4mm}}]

This document supplements the paper ``Time-Based Sampling in Shift-Invariant Spaces.''
Appendix~\ref{app:proofs} contains the proofs of the results stated in the paper, and
Appendix~\ref{app:sims} contains additional simulations. Equations, figures and tables are
numbered consecutively with the paper, and citations refer to the bibliography of the paper.

\appendices
\section{Proofs}
\label{app:proofs}

\begin{proof}[Proof of Theorem~\ref{thm:main}]
Since $x(t)\in\mathcal{V}$ and $S(\omega)$ is supported on
$|\omega|\leq\pi/T$, the signal
$y(t)$ is $2\Omega$-BL with Nyquist rate $T^{-1}$. Condition
\eqref{eq:density_condition} is equivalent to $\Delta t_{\max} < \pi/\Omega$,
so the IF-TEM fires above the Nyquist rate. By~\cite{lazar2004perfect},
the iterative scheme \eqref{eq:iteration} converges to $y(t)$, yielding
$Y(\omega)$ for $|\omega|\leq\pi/T$.

 Since $Y(\omega)=S(\omega)X(\omega)$ and
$X(\omega)=A(e^{j\omega T})H(\omega)$ by \eqref{eq:si_signal}, we have
\begin{equation}
 Y(\omega) = A(e^{j\omega T})\,S(\omega)H(\omega), \qquad |\omega| \leq \frac{\pi}{T}.
 \label{eq:Y_AH}
\end{equation}
Under the assumption $S(\omega)H(\omega)\neq 0$ a.e., the coefficients are recovered as
\begin{equation}
 A(e^{j\omega T}) = \frac{Y(\omega)}{S(\omega)H(\omega)},
 \qquad |\omega| \leq \frac{\pi}{T}.
 \label{eq:A_recovery}
\end{equation}
Inverting the continuous-time spectrum uses $1/(S(\omega)H(\omega))$. Equivalently, the DTFT of the Nyquist samples $y[nT]$ is $Y(\omega)/T$, so the digital filter applied to these samples has response $T/(S(\omega)H(\omega))$ and extracts $a[n]$, from which $x(t)$ is reconstructed via
\eqref{eq:si_signal}. This is the classical SI correction of
Section~\ref{subsec:si}: for the ideal-LPF prefilter,
$R_{SH}(e^{j\omega T})=H(\omega)/T$ on the band, so $T/H(\omega)$ coincides with the correction filter
$1/R_{SH}(e^{j\omega T})$ of uniform SI sampling, applied here to samples
obtained from the firing times rather than from a clocked sampler. The final
claim of the theorem is immediate from \eqref{eq:firing_rate_limits}.
\end{proof}

\begin{proof}[Proof of Proposition~\ref{prop:stability}]
From \eqref{eq:A_recovery},
 $\hat{A}(e^{j\omega T}) - A(e^{j\omega T})
= (\hat{Y}(\omega)-Y(\omega))/(S(\omega)H(\omega))$
for $|\omega|\leq\pi/T$.
 By the DTFT Parseval identity and
$|S(\omega)H(\omega)|^2 \geq T\alpha_{\mathrm{b}}$ a.e.,
\begin{align*}
\|\hat{a}-a\|_{\ell^2}^2
&= \frac{T}{2\pi}\int_{-\pi/T}^{\pi/T}
 \frac{|\hat{Y}(\omega)-Y(\omega)|^2}{ |S(\omega)H(\omega)|^2}\,d\omega \\
&\leq \frac{T}{2\pi} \cdot \frac{1}{T\alpha_{\mathrm{b}}}
 \int_{-\pi/T}^{\pi/T}|\hat{Y}(\omega)-Y(\omega)|^2\,d\omega \\
&= \frac{1}{\alpha_{\mathrm{b}}} \cdot \frac{1}{2\pi}
 \int_{-\pi/T}^{\pi/T}|\hat{Y}(\omega)-Y(\omega)|^2\,d\omega \\
&= \frac{1}{\alpha_{\mathrm{b}}}\,\|\hat{y}-y\|_{L^2}^2
\leq \frac{1}{\alpha_{\mathrm{b}}}\,\varepsilon^2,
\end{align*}
where the last equality uses the continuous-time Parseval identity
$\|f\|_{L^2}^2 = \frac{1}{2\pi}\int|F(\omega)|^2\,d\omega$.
This gives \eqref{eq:si_error}. Substituting
$\varepsilon=r^{l+1}\|y\|_{L^2}$ from \eqref{eq:convergence} gives
\eqref{eq:si_error_iter}.
\end{proof}

\begin{proof}[Proof of Theorem~\ref{thm:robust}]
Multiplying $x(t)\in\mathcal{V}$ by $p(t)=\sum_l c_l e^{j2\pi lt/T}$ gives,
by the modulation property of the CTFT~\cite{eldar2015sampling},
\begin{equation}
 \mathcal{F}\{p(t)x(t)\}(\omega)
 = \sum_{l\in\mathbb{Z}} c_l\,X\!\left(\omega-\frac{2\pi l}{T}\right).
 \label{eq:mixer_ctft}
\end{equation}
Using $X(\omega)=A(e^{j\omega T})H(\omega)$ and the $T$-periodicity of
$A(e^{j\omega T})$,

\begin{equation}
\begin{aligned}
 \mathcal{F}\{p(t)x(t)\}(\omega)
 = A(e^{j\omega T})\sum_{l\in\mathbb{Z}} c_l\,H\!\left(\omega-\frac{2\pi l}{T}\right)
 \\= A(e^{j\omega T})\,R(\omega).
 \label{eq:mixer_spectrum}
\end{aligned}
\end{equation}

After the LPF,
\begin{equation}
 Y(\omega) = A(e^{j\omega T})\,R(\omega), \qquad |\omega|\leq\frac{\pi}{T}.
 \label{eq:Y_AR}
\end{equation}
Since $y(t)$ is $2\Omega$-BL and \eqref{eq:density_condition}
ensures the IF-TEM fires above the Nyquist rate, the iterative scheme
\eqref{eq:iteration} recovers $y(t)$ and hence $Y(\omega)$. Under
$R(\omega)\neq 0$ a.e.,
\begin{equation}
 A(e^{j\omega T}) = \frac{Y(\omega)}{R(\omega)},
 \qquad |\omega|\leq\frac{\pi}{T}.
 \label{eq:A_recovery_robust}
\end{equation}
The spectral inverse is $1/R(\omega)$; equivalently, a digital correction filter $T/R(\omega)$ applied to the Nyquist samples $\{y[nT]\}$ extracts $a[n]$, from which $x(t)$ is recovered via \eqref{eq:si_signal}.
\end{proof}

\begin{proof}[Proof of Corollary~\ref{cor:riesz}]
Since $h$ is real, $G(\omega)=\overline{H(\omega)}$, and with $c_k=1$ for
$k\in\mathcal{K}$ and $c_k=0$ otherwise, \eqref{eq:R_GP} gives
$R_{GP}(\omega)=R_{\mathcal{K}}(\omega)$; the mixer is real because
$\mathcal{K}=-\mathcal{K}$. The prefiltered signal $(g*x)(t)$ lies in the SI
space generated by $g*h$, so Theorem~\ref{thm:robust} applied to that space,
with $R(\omega)$ replaced by $R_{GP}(\omega)$, gives perfect recovery whenever
$R_{\mathcal{K}}(\omega)\neq0$ a.e.\ on the band, which
$\alpha_{\mathcal{K}}>0$ ensures. For the bound,
$\hat{A}(e^{j\omega T})-A(e^{j\omega T})
=(\hat{Y}(\omega)-Y(\omega))/R_{\mathcal{K}}(\omega)$ on the band and
$R_{\mathcal{K}}(\omega)\geq T\alpha_{\mathcal{K}}$ a.e., so the two Parseval
identities used in the proof of Proposition~\ref{prop:stability} give
\begin{equation*}
 \|\hat{a}-a\|_{\ell^2}^2
 =\frac{T}{2\pi}\int_{-\pi/T}^{\pi/T}
 \frac{|\hat{Y}(\omega)-Y(\omega)|^2}{R_{\mathcal{K}}(\omega)^2}\,d\omega
 \leq\frac{1}{T\alpha_{\mathcal{K}}^2}\,\|\hat{y}-y\|_{L^2}^2,
\end{equation*}
which is \eqref{eq:riesz_stability}. Finally, the partial sums
$R_{\{-m,\ldots,m\}}(\omega)$ are continuous, nondecreasing in $m$, and converge
pointwise to the continuous function $T R_{HH}(e^{j\omega T})$; by Dini's
theorem the convergence is uniform on the compact band, so their minima converge
to $T\min_\omega R_{HH}(e^{j\omega T})=T\alpha_0$.
\end{proof}

\begin{proof}[Proof of Theorem~\ref{thm:multi}]
In each channel $i$, the Fourier transform of $p_i(t)x(t)$ is
\begin{equation}
 \mathcal{F}\{p_i(t)x(t)\}(\omega)
 = \sum_{k\in\mathbb{Z}} c_{i,k}\, X\!\left(\omega-\frac{2\pi k}{T}\right).
 \label{eq:multi_ctft}
\end{equation}
Using $X(\omega) = \sum_{l=1}^N A_l(e^{j\omega T}) H_l(\omega)$ and the
$T$-periodicity of each $A_l(e^{j\omega T})$,
\begin{align}
 \mathcal{F}\{p_i(t)x(t)\}(\omega)
 &= \sum_{k\in\mathbb{Z}} c_{i,k}
 \sum_{l=1}^N A_l(e^{j\omega T})\, H_l\!\left(\omega-\frac{2\pi k}{T}\right)
 \nonumber\\
 &= \sum_{l=1}^N A_l(e^{j\omega T})
 \sum_{k\in\mathbb{Z}} c_{i,k}\, H_l\!\left(\omega-\frac{2\pi k}{T}\right).
 \label{eq:multi_spectrum}
\end{align}
After the LPF in each channel,
\begin{equation}
 Y_i(\omega) = \sum_{l=1}^N A_l(e^{j\omega T})
 \left[\mathbf{M}(\omega)\right]_{i,l},
 \qquad |\omega|\leq\frac{\pi}{T}.
 \label{eq:multi_Y}
\end{equation}
Stacking over all $I$ channels, this reads
\begin{equation}
 \mathbf{Y}(\omega) = \mathbf{M}(\omega)\,\mathbf{A}(e^{j\omega T}),
 \qquad |\omega|\leq\frac{\pi}{T},
 \label{eq:multi_system}
\end{equation}
where $\mathbf{Y}(\omega) = [Y_1(\omega),\ldots,Y_I(\omega)]^\top$ and
$\mathbf{A}(e^{j\omega T}) = [A_1(e^{j\omega T}),\ldots,A_N(e^{j\omega T})]^\top$.

In each channel $i$, $y_i(t)$ is $2\Omega$-BL and the density
condition \eqref{eq:density_condition} ensures the IF-TEM fires above the
Nyquist rate, so $y_i(t)$ is recovered from $\{t_n^{(i)}\}$ via the
iterative scheme \eqref{eq:iteration}, yielding $Y_i(\omega)$ for
$|\omega|\leq\pi/T$. Given all $\{Y_i(\omega)\}_{i=1}^I$, the coefficient
vector $\mathbf{A}(e^{j\omega T})$ is recovered by the pseudoinverse
\begin{equation}
 \mathbf{A}(e^{j\omega T}) = \mathbf{M}^\dagger(\omega)\,\mathbf{Y}(\omega),
 \qquad |\omega|\leq\frac{\pi}{T},
 \label{eq:multi_recovery}
\end{equation}
which is well defined almost everywhere since $\mathbf{M}(\omega)$ has rank at
least $N$ a.e. Each $a_l[n]$ is then obtained by inverse DTFT of
$A_l(e^{j\omega T})$.
\end{proof}

\begin{proof}[Proof of Proposition~\ref{prop:stability_multi}]
From \eqref{eq:multi_recovery}, the error in the recovered coefficients is
\begin{equation}
 \hat{\mathbf{A}}(e^{j\omega T}) - \mathbf{A}(e^{j\omega T})
 = \mathbf{M}^\dagger(\omega)
 \bigl(\hat{\mathbf{Y}}(\omega) - \mathbf{Y}(\omega)\bigr).
 \label{eq:multi_err_freq}
\end{equation}
By assumption \eqref{eq:sigma_lower},
$\|\mathbf{M}^\dagger(\omega)\|_2 = 1/\sigma_{\min}(\omega) \leq
1/(\sqrt{T}\,\sigma)$ a.e.
The DTFT Parseval identity gives
\begin{align}
 \sum_{l=1}^N \|\hat{a}_l - a_l\|_{\ell^2}^2
 &= \frac{T}{2\pi}\int_{-\pi/T}^{\pi/T}
 \bigl\|\mathbf{M}^\dagger(\omega)
 \bigl(\hat{\mathbf{Y}}-\mathbf{Y}\bigr)\bigr\|_2^2\,d\omega \nonumber\\
 &\leq \frac{T}{2\pi} \cdot \frac{1}{T\sigma^2}
 \int_{-\pi/T}^{\pi/T}
 \|\hat{\mathbf{Y}}(\omega)-\mathbf{Y}(\omega)\|_2^2\,d\omega \nonumber\\
 &= \frac{1}{\sigma^2} \cdot \frac{1}{2\pi}
 \int_{-\pi/T}^{\pi/T}
 \|\hat{\mathbf{Y}}(\omega)-\mathbf{Y}(\omega)\|_2^2\,d\omega \nonumber\\
 &= \frac{1}{\sigma^2}
 \sum_{i=1}^I \|\hat{y}_i - y_i\|_{L^2}^2,
 \label{eq:multi_err_bound}
\end{align}
where the last step uses the continuous-time Parseval identity
$\|f\|_{L^2}^2 = \frac{1}{2\pi}\int|F(\omega)|^2\,d\omega$ applied
channel-wise. The final inequality in \eqref{eq:multi_stability} follows
from $\|\hat{y}_i - y_i\|_{L^2} \leq \varepsilon$ for all $i$.
\end{proof}

\begin{proof}[Proof of Lemma~\ref{lem:bl_mse}]
This follows directly from~\cite[Theorem~2]{naaman_quant} applied to $y(t)$
with bandwidth $\Omega = \pi/T$ and energy $E_y$, using the choice
$c = \sqrt{E_y/T}$ of \eqref{eq:c_Ey} and $b = \alpha c$. The condition
$R < 1$ is equivalent to the density condition \eqref{eq:density_condition}
under the parametrization $b = \alpha c$.
\end{proof}

\begin{proof}[Proof of Theorem~\ref{thm:quant_si}]
Let $P$ be the orthogonal projection onto signals supported on
$|\omega|\leq\pi/T$. Since $Py=y$, Proposition~\ref{prop:stability} applied to
$P(\mathbf{1}_{W_n}\hat{y})$, together with
$\|P(\mathbf{1}_{W_n}\hat{y}-y)\|_{L^2}\leq\|\mathbf{1}_{W_n}\hat{y}-y\|_{L^2}$, gives
\begin{align}
 \sum_{k\in J_n}|\hat{a}_n[k]-a[k]|^2
 &\leq \frac{1}{\alpha_{\mathrm{b}}}\,\|\mathbf{1}_{W_n}\hat{y}-y\|_{L^2}^2\nonumber\\
 &=\frac{1}{\alpha_{\mathrm{b}}}\bigl(\|\mathbf{1}_{W_n}e\|_{L^2}^2
 +\|\mathbf{1}_{W_n^c}\,y\|_{L^2}^2\bigr).
 \label{eq:si_prop_applied}
\end{align}
Divide by $M_n$ and let $n\to\infty$. Since $|W_n|/M_n\to T$ and $y\in L^2$, the
second term vanishes, and by \eqref{eq:mse_def} the upper limit is at most
$(T/\alpha_{\mathrm{b}})\,\varepsilon^2$. Taking expectations gives the first
inequality in \eqref{eq:si_coeff_mse}, and Lemma~\ref{lem:bl_mse} gives the
second. For \eqref{eq:si_coeff_mse_x}, write $u=\sqrt{E_y}$ and
$c_0=\kappa\delta/((\alpha-1)\sqrt{T})$, so that $R=c_0/u$. The
$E_y$-dependent factor of \eqref{eq:si_coeff_mse} is
$g(u)=c_0\,u^3/(u-c_0)^2$, whose derivative
$g'(u)=c_0\,u^2(u-3c_0)/(u-c_0)^3$ is nonnegative for $u\geq 3c_0$, that is,
for $R\leq\frac13$. Since $E_y\leq\beta_{\mathrm{b}}\|a\|_{\ell^2}^2$ by
\eqref{eq:Ey_bounds}, $g$ is nondecreasing between $\sqrt{E_y}$ and
$\|a\|_{\ell^2}\sqrt{\beta_{\mathrm{b}}}$, and replacing $E_y$ by its upper
bound yields \eqref{eq:si_coeff_mse_x}.
\end{proof}

\section{Additional Simulations}
\label{app:sims}

\subsection{Noise-Free Recovery Across Signal Classes}
\label{subsec:sim_recovery}

\begin{figure*}[t!]
\centering
\begin{subfigure}{0.42\linewidth}
 \includegraphics[width=\linewidth]{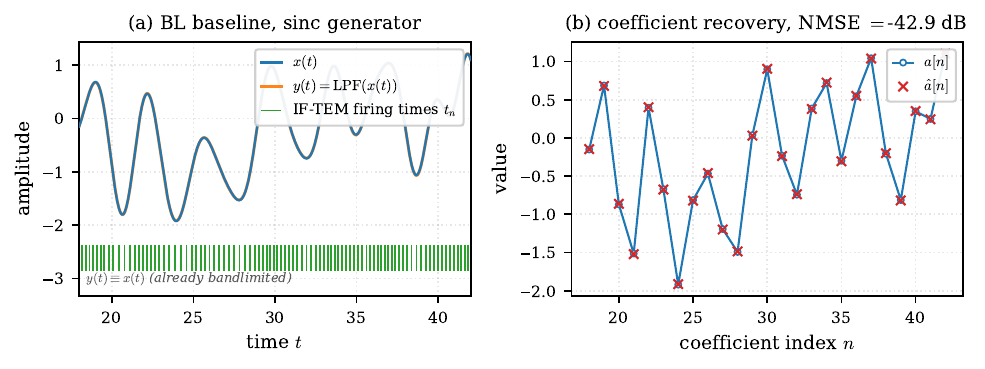}
 \caption{BL baseline (sinc generator).}
 \label{fig:rec_bl}
\end{subfigure}\hfill
\begin{subfigure}{0.42\linewidth}
 \includegraphics[width=\linewidth]{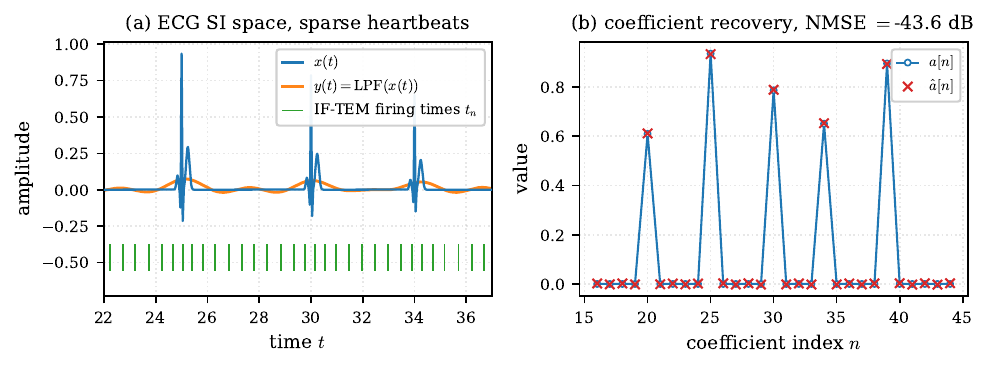}
 \caption{ECG generator (real heartbeat pulse).}
 \label{fig:rec_ecg}
\end{subfigure}

\vspace{2mm}

\begin{subfigure}{0.42\linewidth}
 \includegraphics[width=\linewidth]{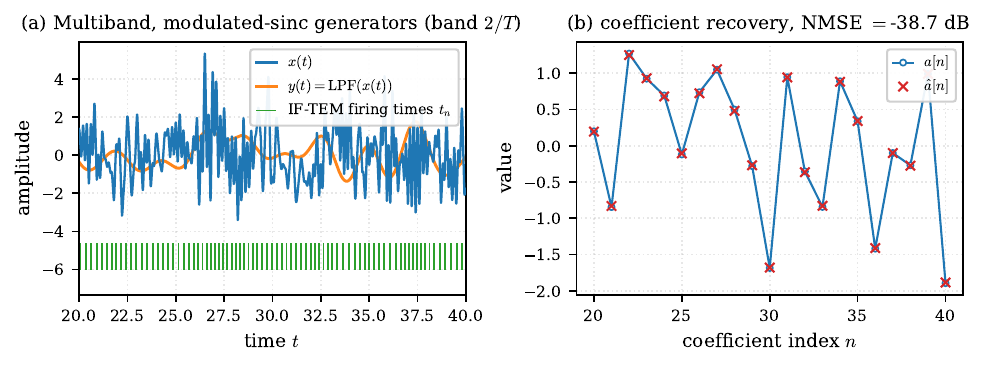}
 \caption{Multiband (modulated-sinc generators).}
 \label{fig:rec_multiband}
\end{subfigure}\hfill
\begin{subfigure}{0.42\linewidth}
 \includegraphics[width=\linewidth]{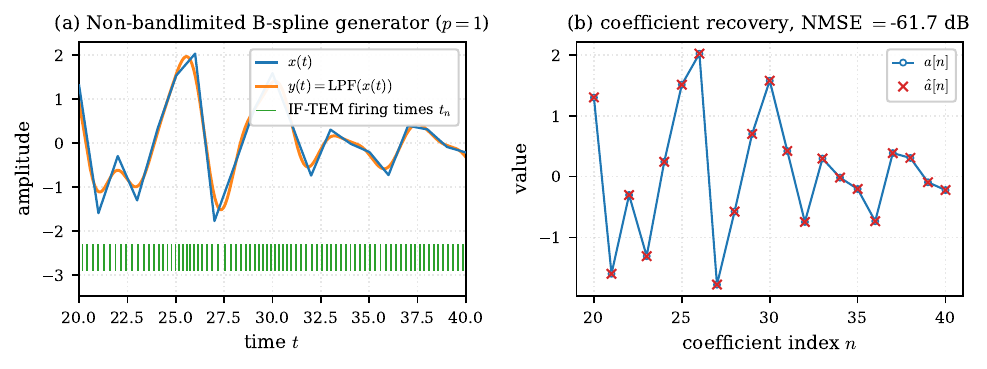}
 \caption{Non-BL B-spline generator.}
 \label{fig:rec_splines}
\end{subfigure}
\caption{Noise-free IF-TEM sampling and recovery of SI signals for four signal
classes. In each subfigure, panel (a) shows the input $x(t)$ (blue), its
low-pass version $y(t)=\mathrm{LPF}_{\pi/T}\{x(t)\}$ (orange), and the IF-TEM
firing times $\{t_n\}$ (green); panel (b) compares the true coefficients
$a[n]$ with the reconstruction $\hat{a}[n]$. The firing-time density visibly
adapts to local signal activity, most clearly for the ECG generator.}
\label{fig:recovery}
\end{figure*}

We illustrate the complete pipeline on four signal classes, each highlighting a
different aspect of the framework, following the format of
\cite[Fig.~5]{kvich2024modulo}: panel (a) of each subfigure in
Fig.~\ref{fig:recovery} overlays the input $x(t)$, its low-pass version $y(t)$
and the firing times $\{t_n\}$, while panel (b) compares $a[n]$ with
$\hat{a}[n]$.

\emph{(i) BL baseline.} With $h(t)=\mathrm{sinc}(\pi t/T)$ the SI space
coincides with the $2\Omega$-BL signals and $x(t)\equiv y(t)$. This canonical
IF-TEM setting serves as a sanity check and recovers $a[n]$ with a coefficient NMSE of $-42.9$~dB.

\emph{(ii) Electrocardiogram (ECG).} A single measured heartbeat is the
generator, and the SI signal is a stream of heartbeats on the grid. We
deliberately operate at a low firing rate ($\mathrm{OF}=2$), so that the firing
times are sparse and strongly non-uniform: the inter-spike intervals tighten
across each QRS complex and stretch over the flat baseline. This
signal-adaptive, clock-free behavior is the defining feature of the IF-TEM, and
the coefficients are still recovered accurately from the reduced sample set.

\emph{(iii) Multiband.} The space uses $N$ modulated-sinc generators centered at
distinct harmonics of $1/T$, recovered from $I=N$ mixer channels via
$\mathbf{M}^\dagger(\omega)$. Although the highest frequency present is many
times $1/T$, the per-channel firing rate stays tied to $T^{-1}$, so acquisition
is strongly sub-Nyquist with respect to the maximum frequency: it is the in-band
energy, and not the bandwidth, that governs the IF-TEM operating point.

\emph{(iv) Splines.} The generator is a B-spline, which is \emph{not} BL, so
$x(t)$ has energy beyond $\pi/T$ and the LPF visibly smooths the
piecewise-polynomial input. The coefficients are extracted by the correction
filter of Theorem~\ref{thm:main}, giving accurate recovery of a non-BL signal at a firing rate set by the SI coefficient rate rather than by the generator bandwidth.

\subsection{Noise Resilience without In-Band Zeros}
\label{subsec:sim_noise}

\begin{figure}[t!]
 \centering
 \includegraphics[width=0.6\linewidth]{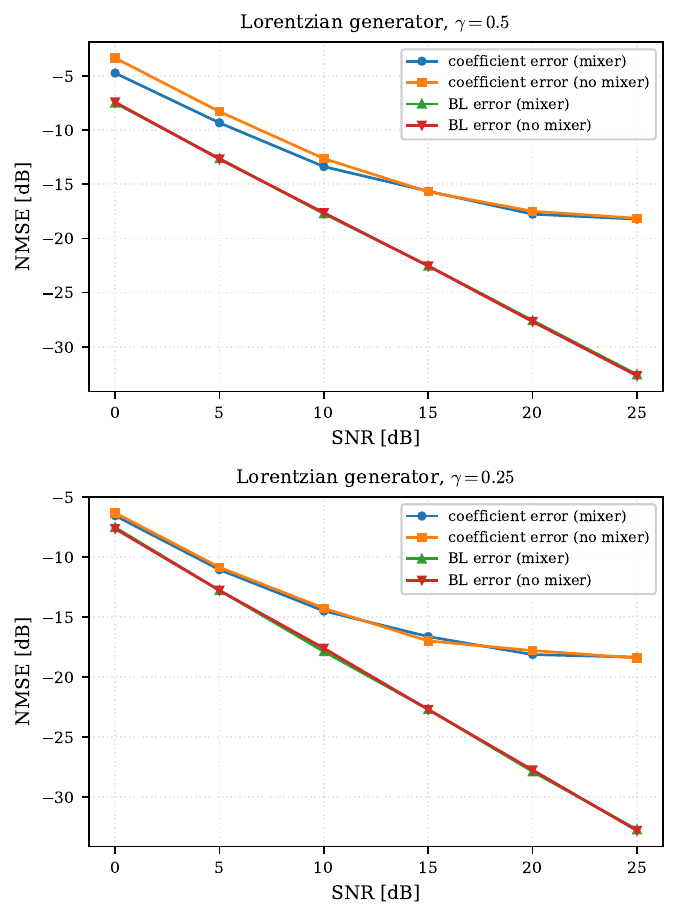}
 \caption{NMSE in dB versus SNR for a Lorentzian generator with scales
 $\gamma=0.5$ (top) and $\gamma=0.25$ (bottom). ``BL error'' refers to the
 recovery discrepancy of $\hat{y}[nT_s]$ in \eqref{eq:nmse_bl}, and
 ``coefficient error'' to the precision of $\hat{a}[n]$ in
 \eqref{eq:nmse_coef}, each shown with and without the mixer.
 The coefficient-error curves with and without mixing nearly
 coincide, whereas their separation from the BL-error curves increases at
 higher SNR.}
 \label{fig:snr_lorentzian}
\end{figure}

We next study robustness to noise for a generator whose
baseband response is bounded away from zero. Following
\cite{kvich2024modulo}, we use the Lorentzian generator $h_\gamma(t) =
\tfrac{1}{\pi\gamma}\bigl(1+(t/\gamma)^2\bigr)^{-1}$ with CTFT
$H_\gamma(\omega) = e^{-\gamma|\omega|}$, which is bounded away from zero on
$|\omega|\leq\pi/T$ and therefore yields a well-conditioned correction filter
$1/H_\gamma(\omega)$. Both NMSE metrics are reported versus SNR in
Fig.~\ref{fig:snr_lorentzian}, for the two scale parameters $\gamma = 0.5$ and
$\gamma=0.25$, with and without a mixer. The OF is fixed at $5$.

Two observations stand out. First, the coefficient-error curves
with and without mixing differ by at most approximately $1.5$~dB over the
plotted SNR range. At higher SNR, the coefficient NMSE approaches $-18$~dB,
while the BL error continues to decrease. Thus, mixing provides little
improvement for these generators, whose baseband responses have no zeros. Second, the LPF stage incurs a generator- and
scale-dependent energy loss, which enters the quantization step
\eqref{eq:quant_step_Ey} and the bounds \eqref{eq:Ey_bounds} through $E_y$. The
narrower-band $\gamma=0.5$ generator loses only a few percent of its energy to
the LPF, whereas the wider-band $\gamma=0.25$ generator loses substantially
more. This confirms the role of $E_y$ as the energy that survives the LPF and
drives the IF-TEM operating point, consistent with \eqref{eq:Ey_bounds}.

\subsection{Effective and Correction Filters for the Scaled B-Spline}
\label{app:corr_filter}

\begin{figure}[t!]
 \centering
 \includegraphics[width=0.5\linewidth]{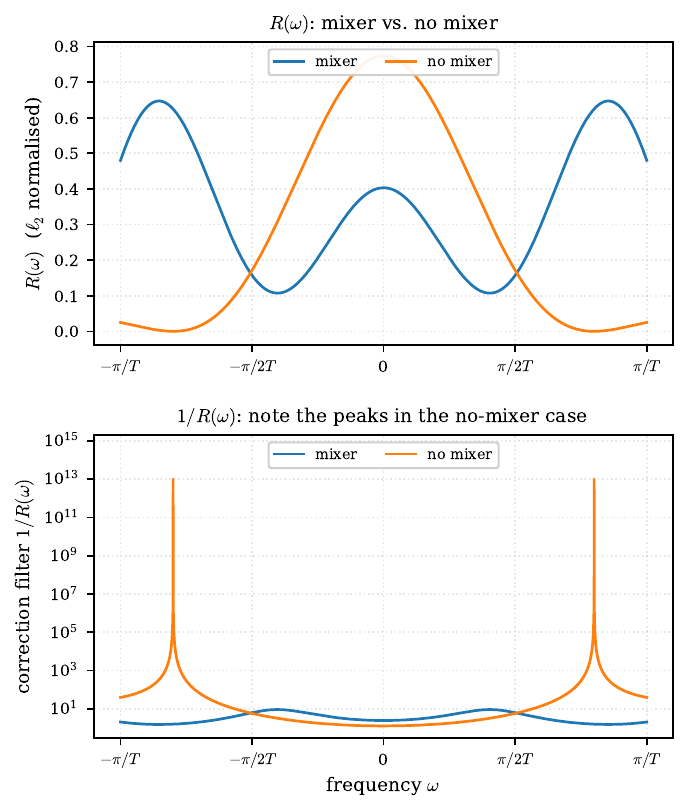}
 \caption{Top: the effective filter $R(\omega)$ obtained in the simulations
 for the mixer and non-mixer scenarios, after $\ell_2$ normalization. Bottom:
 the associated correction filter $1/R(\omega)$. In the non-mixer case
 $R(\omega)$ has in-band zeros, producing the sharp peaks in $1/R(\omega)$;
 the mixer removes the zeros and the associated singular
 peaks in the correction filter.}
 \label{fig:corr_filter}
\end{figure}

The mechanism behind the behavior in Fig.~\ref{fig:snr_spline} is the shape of the effective filter
$R(\omega)$ and of the correction filter $1/R(\omega)$, plotted for both
scenarios in Fig.~\ref{fig:corr_filter} after $\ell_2$ normalization. Without a
mixer, $R(\omega)=H(\omega)$ vanishes inside the band, producing sharp peaks in
$1/R(\omega)$ that amplify any noise at those frequencies. With the mixer, $R(\omega)$ is bounded away from zero,
so the correction filter no longer has the singular peaks of the unmixed
response. This is precisely the conditioning improvement
described after Theorem~\ref{thm:robust} and quantified through the lower bound
on $|R(\omega)|^2/T$ in Proposition~\ref{prop:stability}.

\end{document}